\documentclass[review]{elsarticle}
\usepackage{fullpage}

\usepackage{hyperref}
\usepackage{amsmath}
\usepackage{amssymb}
\usepackage{tabls}
\usepackage{algorithm}

\journal{International Journal of Approximate Reasoning}

\newcommand{\Score}{\mathit{Score}}
\newcommand{\Test}{\mathit{Test}}
\newcommand{\mref}[1]{(\ref{#1})}

\newcommand{\ck}{\checkmark}
\newcommand{\dd}{\,d}

\newcommand{\Prob}{\operatorname{P}}

\newcommand{\given}{\operatorname{|}}
\newcommand{\X}{\mathbf{X}}
\newcommand{\G}{\mathcal{G}}
\newcommand{\D}{\mathcal{D}}
\newcommand{\TO}{\mathcal{T}}

\newcommand{\PXi}{\Pi_{X_i}}
\newcommand{\XPi}{X_i \given \PXi}
\newcommand{\T}{\Theta_{X_i}}
\newcommand{\TT}{\T \given \PXi}

\newcommand{\BG}{\operatorname{BG}}
\newcommand{\xm}{\overline{\mathbf{x}}}
\newcommand{\am}{\alpha_{\mu}}
\newcommand{\aw}{\alpha_w}
\newcommand{\nnu}{\boldsymbol{\nu}}
\newcommand{\BD}{\operatorname{BD}}

\newcommand{\aijk}{\alpha_{ijk}}
\newcommand{\piijk}{\pi_{ik \given j}}

\newcommand{\Gp}{\mathcal{G}^{+}}
\newcommand{\Gm}{\mathcal{G}^{-}}
\newcommand{\Gmax}{\G_{\mathit{max}}}
\renewcommand{\S}{\mathbf{S}}
\newcommand{\Smax}{S_{\mathit{max}}}
\newcommand{\Z}{\mathbf{Z}}
\newcommand{\BIC}{\mathrm{BIC}}
\newcommand{\BICg}{\BIC_{\gamma}}
\newcommand{\GSQ}{\mathrm{G}^2}
\newcommand{\GB}{\GSQ_\BIC}
\newcommand{\BF}{\mathrm{BF}}
\newcommand{\XT}{\mathrm{X}^2}

\newcommand{\rXYZ}{\rho_{XY \given \Z}}

\newcommand{\HHPC}{H\textsuperscript{2}PC}

\begin{document}

\begin{frontmatter}

\title{Who Learns Better Bayesian Network Structures: Accuracy and Speed of
    Structure Learning Algorithms}

\author{Marco Scutari}
\ead{scutari@idsia.ch}
\address{Istituto Dalle Molle di Studi sull'Intelligenza Artificiale (IDSIA),
  Lugano, Switzerland}

\author{Catharina Elisabeth Graafland}
\author{Jos{\'e} Manuel Guti{\'e}rrez}
\address{Institute of Physics of Cantabria (CSIC-UC), Santander, Spain}

% Abstract and keywords
\begin{abstract}%
  Three classes of algorithms to learn the structure of Bayesian networks from
  data are common in the literature: \emph{constraint-based algorithms}, which
  use conditional independence tests to learn the dependence structure of the
  data; \emph{score-based algorithms}, which use goodness-of-fit scores as
  objective functions to maximise; and \emph{hybrid algorithms} that combine
  both approaches. Constraint-based and score-based algorithms have been shown
  to learn the same structures when conditional independence and goodness of fit
  are both assessed using entropy and the topological ordering of the network is
  known \cite{cowell}.

  In this paper, we investigate how these three classes of algorithms perform
  outside the assumptions above in terms of speed and accuracy of network
  reconstruction for both discrete and Gaussian Bayesian networks. We approach
  this question by recognising that structure learning is defined by the
  combination of a \emph{statistical criterion} and an \emph{algorithm} that
  determines how the criterion is applied to the data.  Removing the confounding
  effect of different choices for the statistical criterion, we find using both
  simulated and real-world complex data that constraint-based algorithms are
  often less accurate than score-based algorithms, but are seldom faster (even
  at large sample sizes); and that hybrid algorithms are neither faster nor more
  accurate than constraint-based algorithms. This suggests that commonly held
  beliefs on structure learning in the literature are strongly influenced by the
  choice of particular statistical criteria rather than just by the properties
  of the algorithms themselves.

\end{abstract}

\end{frontmatter}

\section{Background and Notation}
\label{sec:bg}

Bayesian networks \cite[BNs; ][]{koller} are a class of graphical models
defined over a set of random variables $\X = \{X_1, \ldots, X_N\}$, each
describing some quantity of interest, that are associated with the nodes of a
directed acyclic graph (DAG) $\G$. (They are often referred to interchangeably.)
The structure of the DAG, that is, the set of arcs $A$ of $\G$, encodes the
independence relationships between those variables, with graphical separation
in $\G$ implying conditional independence in probability. As a result, $\G$
induces the factorisation
\begin{equation}
  \label{eq:parents}
  \Prob(\X \given \G, \Theta) = \prod_{i=1}^N \Prob(\XPi, \T),
\end{equation}
in which the \emph{global distribution} of $\X$ (with parameters $\Theta$)
decomposes in one \emph{local distribution} for each $X_i$ (with parameters
$\T$) conditional on its parents $\PXi$. This decomposition holds only in the
absence of missing data, which we will assume in the following.

The DAG $\G$ does not uniquely identify a single BN: all BNs with the same
underlying undirected graph and v-structures (patterns of arcs like $X_i
\rightarrow X_j \leftarrow X_k$, with no arc between $X_i$ and $X_k$) fall into
the same \emph{equivalence class} \cite{chickering} of models and are
probabilistically indistinguishable. It is easy to see that the two other
possible patterns of two arcs and three nodes result in equivalent
factorisations:
\begin{equation}
  \underbrace{\Prob(X_i)\Prob(X_j \given X_i)\Prob(X_k \given X_j)}_{X_i \rightarrow X_j \rightarrow X_k} =
  \Prob(X_j, X_i)\Prob(X_k \given X_j) =
  \underbrace{\Prob(X_i \given X_j)\Prob(X_j)\Prob(X_k \given X_j)}_{X_i \leftarrow X_j \rightarrow X_k}.
\end{equation}
Hence each equivalence class is represented by the completed partially-directed
acyclic graph (CPDAG) that arises from the combination of these two quantities.

While in principle there are many possible choices for the distribution of $\X$,
the literature has focused for the most part on two sets of assumptions:
\begin{itemize}
  \item \emph{Discrete BNs} \cite{heckerman} assume that the $X_i$ are
    multinomial random variables:
    \begin{equation*}
      \XPi \sim \mathit{Mul}(\piijk), \qquad \piijk = \Prob(X_i = k \given \PXi = j);
    \end{equation*}
    the $\piijk$ are the conditional probabilities of $X_i$ given the $j$th
    configuration of the values of its parents. As a result, $\X$ is also
    multinomial. When learning BNs from data, generally we further assume
    positivity ($\piijk > 0$), parameter independence ($\piijk$ for different
    parent configurations are independent) and parameter modularity ($\piijk$
    associated with different nodes are independent).
  \item \emph{Gaussian BNs} \cite[GBNs;][]{heckerman3} assume that the $X_i$
    are univariate normal random variables linked by linear dependencies to
    their parents,
    \begin{equation*}
      \XPi \sim N(\mu_{X_i} + \PXi\boldsymbol{\beta}_{X_i}, \sigma^2_{X_i}),
    \end{equation*}
    in what is essentially a linear regression model of $X_i$ against the $\PXi$
    with regression coefficients $\boldsymbol{\beta}_{X_i} =\{ \beta_{X_i,X_j},
    X_j \in \PXi\}$. $\X$ is then multivariate normal, and we generally assume
    that its covariance matrix $\Sigma$ is positive definite. Equivalently
    \cite{weatherburn}, we can consider the precision matrix $\Omega =
    \Sigma^{-1}$ and parameterise the $\XPi$ with the partial correlations
    \begin{align*}
      &\rho_{X_i, X_j \given \PXi \setminus X_j} =
         \frac{\Omega_{ij}}{\sqrt{\Omega_{ii}\Omega_{jj}}}
    \end{align*}
    between $X_i$ and each parent $X_j \in \PXi$ given the rest, since
    \begin{equation*}
      \beta_{X_i, X_j} = \rho_{X_i, X_j \given \PXi \setminus X_j}
                         \sqrt{\frac{\Omega_{ii}}{\Omega_{jj}}}.
    \end{equation*}
\end{itemize}

Other distributional assumptions have seen less widespread adoption for various
reasons. For instance, copulas \cite{copula} and truncated exponentials
\cite{truncexp} lack exact conditional inference and simple closed-form
estimators; and conditional linear Gaussian BNs \cite{lauritzen} cannot encode
DAGs with arcs pointing from discrete to continuous nodes.

\section{Learning a Bayesian Network from Data}
\label{sec:intro-learning}

The task of learning a BN with DAG $\G$ and parameters $\Theta$ from a data
set $\D$ containing $n$ observations can be performed in two steps in an
inherently Bayesian fashion:
\begin{equation}
  \underbrace{\Prob(\G, \Theta \given \D)}_{\text{learning}} =
    \underbrace{\Prob(\G \given \D)}_{\text{structure learning}} \cdot
    \underbrace{\Prob(\Theta \given \G, \D)}_{\text{parameter learning}}.
\label{eq:lproc}
\end{equation}
\emph{Structure learning} consists in finding the DAG $\G$ that encodes the
dependence structure of the data; \emph{parameter learning} consists
in estimating the parameters $\Theta$ given the $\G$ obtained from structure
learning. If we assume parameters in different local distributions are
independent, they can be learned separately and efficiently for each node
because \mref{eq:parents} then implies
\begin{equation*}
  \Prob(\Theta \given \G, \D) = \prod_{i=1}^N \Prob(\TT, \D).
\end{equation*}
On the other hand, structure learning is well known to be NP-complete
\cite{npcomp}, even when assuming the availability of an independence and
inference oracle \cite{nplarge}; only some relaxations such as \cite{notnphard}
are not NP-hard. Using Bayes theorem once more, we can formulate it as
\begin{equation*}
  \Prob(\G \given \D) \propto \Prob(\G)\Prob(\D \given \G);
\label{eq:marginal-likelihood}
\end{equation*}
and following \mref{eq:parents} we can decompose the marginal likelihood
$\Prob(\D \given \G)$ into one component for each local distribution
\begin{equation}
  \Prob(\D \given \G)
    = \int \Prob(\D \given \G, \Theta) \Prob(\Theta \given \G) \dd\Theta = \\
    = \prod_{i=1}^N \int \Prob(\XPi, \T) \Prob(\TT) \dd\T.
\label{eq:local-score}
\end{equation}
Closed-form expressions for \mref{eq:local-score} are available for both
discrete BNs and GBNs; and \mref{eq:local-score} can be approximated using the
Bayesian information criterion (BIC) \cite{schwarz} as well. Both will be
described in Section \ref{sec:intro-criterion}. As for $\Prob(\G)$, the most
common choice in the literature is the uniform distribution; we will default to
it in the following as well. The space of the DAGs grows super-exponentially in
$N$ \citep{harary} and that makes it cumbersome to specify informative priors:
two notable exceptions are presented in \cite{csprior} and \cite{mukherjee}.
\cite{csprior} described a \textit{completed prior} in which they elicitated
prior probabilities for a subset of arcs and completed the prior to cover the
remaining arcs with a discrete uniform distribution. As an alternative,
\cite{mukherjee} proposed an informative prior using a log-linear combination of
arbitrary patterns of arcs. Some structure learning approaches
\cite[\emph{e.g.}][]{k2} also assume the topological ordering of $\G$ to be
known \emph{a priori} and assign a prior probability of zero to any DAG that is
incompatible with that ordering. This effectively assigns a prior probability of
zero to many arcs; and it completely side-steps the identifiability issues
arising from the existence of equivalence classes because, for each arc, only
one direction is compatible with the topological ordering.

\subsection{Structure Learning Algorithms}
\label{sec:intro-algorithms}

Several algorithms have been proposed to implement BN structure learning,
following one of three possible approaches: \emph{constraint-based},
\emph{score-based} and \emph{hybrid}.

\begin{algorithm}[t]
\caption{PC-Stable Algorithm}
\label{algo:pc}

\textbf{Input:} a data set $\D$ from $\X$, a (conditional) independence test
  $\Test(X_i, X_j \given \S; \D)$. \\
\textbf{Output:} a CPDAG $\G$.

\begin{enumerate}
  \item Initialise a complete undirected graph $\G$ spanning $\X$.
  \label{step:pc1}
  \item For $l = 0, 1, \ldots, N - 2$: \label{step:pc2}
    \begin{enumerate}
      \item For all adjacent pairs of nodes $(X_i, X_j), i \neq j$ such that
        $X_i$ has at least $l$ neighbours in the current $\G$, excluding $X_j$:
        \begin{enumerate}
          \item Choose a new subset $\S$ of size $l$ from the neighbours of
            $X_i$ excluding $X_j$;
          \item If $\Test(X_i, X_j \given \S; \D)$ accepts the hypothesis that
            $X_i$ is independent from $X_j$ given $\S$, remove $X_i - X_j$ from
            $\G$ and set $\S_{X_i X_j} = \S$ as the separating set of
            ($X_i, X_j)$.
          \item If $X_i$ and $X_j$ are no longer adjacent or there are no more
            possible subsets $\mathbf{S}$ of size $l$ to consider, move to the
            next pair of nodes.
        \end{enumerate}
    \end{enumerate}
  \item For each triplet $X_i - X_k - X_j$ such that $X_i$ is not adjacent to
    $X_j$ and that $X_k \notin \mathbf{S}_{X_i X_j}$, replace it with the
    v-structure $X_i \rightarrow X_k \leftarrow X_j$. \label{step:pc3}
  \item Set more arc directions by applying recursively the following two rules:
    \label{step:pc4}
    \begin{enumerate}
      \item if $X_i$ is adjacent to $X_j$ and there is a strictly directed path
        from $X_i$ to $X_j$ then replace $X_i - X_j$ with $X_i \rightarrow X_j$
        (to avoid introducing cycles);
      \item if $X_i$ and $X_j$ are not adjacent but $X_i \rightarrow X_k$ and
        $X_k - X_j$, then replace the latter with $X_k \rightarrow X_j$ (to
        avoid introducing new v-structures).
    \end{enumerate}
\end{enumerate}
\end{algorithm}

Constraint-based algorithms are based on the seminal work of Pearl on causal
graphical models \cite{ic}, which found its first practical implementation in
the PC algorithm \cite{spirtes}. Its modern implementation, called
\emph{PC-stable} \cite{colombo}, is illustrated in Algorithm \ref{algo:pc}.
Steps \ref{step:pc1} and \ref{step:pc2} identify which pairs of variables $(X_i,
X_j)$ are connected by an arc, regardless of its direction. Such variables
cannot be separated by any subset of the other variables; this condition is
tested heuristically by performing a sequence of \emph{conditional independence
tests} $\Test(X_i, X_j \given \S; \D)$ with increasingly large candidate
separating sets $\S$. Step \ref{step:pc3} identifies the \mbox{v-structures}
among all the pairs of non-adjacent nodes $X_i$ and $X_j$ with a common
neighbour $X_k$ using the separating sets found in step \ref{step:pc2}. At the
end of step \ref{step:pc3} both the skeleton and the \mbox{v-structures} of the
network are known; step \ref{step:pc4} then sets the remaining arc directions
using the rules from \cite{chickering} to obtain the CPDAG describing the
identified equivalence class. More recent algorithms such as Grow-Shrink (GS)
\cite{mphd} and Inter-IAMB \cite{fastiamb2} proceed along similar lines, but use
faster heuristics to implement the first two steps; an overview can be found in
\cite{hitonpc}.

\begin{algorithm}[t]
\caption{Greedy Search}
\label{algo:greedy}

\textbf{Input:} a data set $\D$ from $\X$, an initial (usually empty) DAG $\G$,
  a score function $\Score(\G, \D)$. \\
\textbf{Output:} the DAG $\Gmax$ that maximises $\Score(\G, \D)$.

\begin{enumerate}
  \item Compute the score of $\G$, $S_{\G} = \Score(\G, \D)$, and set
    $\Smax = S_{\G}$ and $\Gmax = \G$. \label{step:setup}
  \item \textbf{Hill climbing:} repeat as long as $\Smax$ increases:
  \label{step:hc}
  \begin{enumerate}
    \item for every possible arc addition, deletion or reversal in $\Gmax$
      resulting in a DAG:
      \begin{enumerate}
        \item compute the score of the modified DAG $\G^*$,
          $S_{\G^*} = \Score(\G^*, \D)$:
        \item if $S_{\G^*} > \Smax$ and $S_{\G^*} > S_{\G}$, set $\G = \G^*$
          and $S_{\G} = S_{\G^*}$.
      \end{enumerate}
      \item if $S_\G > S_{\mathit{max}}$, set $\Smax = S_\G$ and $\Gmax = \G$.
  \end{enumerate}
  \item \textbf{Tabu list:} for up to $t_0$ times, repeat step \ref{step:hc}
    but choose the DAG $\G$ with the highest $S_\G$ that has not been visited in
    the last $t_1$ steps regardless of $\Smax$. If a DAG such that $S_\G > \Smax$
    is found, restart the search from step \ref{step:hc}. \label{step:tabu}
  \item \textbf{Random restart:} for up to $r$ times, perturb $\Gmax$ with
    multiple arc additions, deletions and reversals to obtain a new DAG $\G'$
    and search from step \ref{step:hc}. \label{step:restart}
\end{enumerate}
\end{algorithm}

Score-based algorithms represent the application of general-purpose optimisation
techniques to BN structure learning. Each candidate DAG is assigned a
\emph{network score} reflecting its goodness of fit, which the algorithm then
attempts to maximise. Some examples are heuristics such as \emph{greedy search},
\emph{simulated annealing} \cite{bouckaert} and \emph{genetic algorithms}
\cite{larranaga}; a comprehensive review of these and other approaches is
provided in \cite{norvig}. They can also be applied to CPDAGs, as in the case of
Greedy Equivalent Search \cite[GES; ][]{ges}. In recent years exact maximisation
of $\Prob(\G \given \D)$ and BIC has become feasible as well for small data sets
thanks to increasingly efficient pruning of the space of the DAGs and tight
bounds on the scores \citep{cutting,suzuki17,scanagatta}. Another possible
choice is exploring the space of DAGs using Markov chain Monte Carlo methods,
which have the advantage of producing a sample of DAGs from $\Prob(\G \given
\D)$ thus making posterior inference possible. This approach, which dates back
to \cite{madigan-york}, has been improved upon \cite{husmeier-mcmc,kuipers-mcmc}
by first sampling from the space of topological orderings to accelerate mixing.

Greedy search, illustrated in Algorithm \ref{algo:greedy}, represents by far the
most common group score-based algorithm in practical applications. It consists
of an initialisation phase (step \ref{step:setup}) followed by a \emph{hill
climbing} search (step \ref{step:hc}), which is then optionally refined with a
\emph{tabu list} (step \ref{step:tabu}) and \emph{random restarts} (step
\ref{step:restart}). In each iteration, hill climbing tries to delete and to
reverse each arc in the current candidate DAG $\Gmax$; and to add each possible
arc that is not already present in $\Gmax$ and that does not introduce any
cycles. These are local moves that impact only one or two local distributions in
th BN, which greatly reduces the computational complexity of greedy search by
avoiding the need to re-score all nodes at each iterations.  The resulting $\G$
with the highest score $S_{\G}$ is compared to $\Gmax$; if it has a better score
($S_{\G} > \Smax$) then $\G$ becomes the new $\Gmax$. If, on the other hand,
$S_{\G} < \Smax$, greedy search has reached an optimum. There is no guarantee
that $\G$ is a global optimum; hence greedy search may perform further steps to
reduce the chances that $\G$ is in fact a sub-optimal local optimum. One option
is to restart the search in step \ref{step:hc} from a different starting point,
obtained by changing $r$ arcs in the current optimal $\G$. This gives what is
called the \emph{hill climbing with random restarts} algorithm. Another option
is to keep a tabu list of previously-visited DAGs and to continue searching for
a better DAG that has yet been considered, giving the \emph{tabu search}
algorithm. Clearly, it is possible to perform both steps \ref{step:tabu} and
\ref{step:restart} and obtain a tabu search with random restarts.

\begin{algorithm}[t]
\caption{A Simulated Annealing Approach to Structure Learning}
\label{algo:sann}

\textbf{Input:} a data set $\D$, and initial node ordering $\TO_0$, a
  score function $\Score(\G, \D)$ \\
\textbf{Output:} the DAG $\Gmax$ that maximises $\Score(\G, \D)$.

\begin{enumerate}
  \item For a large number of iterations i = 1, \ldots, m: \label{step:order}
    \begin{enumerate}
      \item Generate a new topological ordering $\TO_i$ by randomly permuting
        the nodes in $\TO_{i - 1}$.
      \item Accept the new ordering with some probability $\Prob(\TO_i \given
        \TO_{i -1}, \beta)$, where $\beta$ is the temperature; otherwise
        $\TO_i = \TO_{i -1}$.
      \item Reduce the temperature $\beta$.
    \end{enumerate}
  \item For the best ordering $\widehat\TO$, find the $\G$ with the highest
     $\Score(\G, \D \given \widehat\TO)$. \label{step:conditional}
\end{enumerate}
\end{algorithm}

A second group of score-based algorithms seek to speed-up structure learning by
first obtaining a topological ordering $\TO$ for the nodes, and then learning
the optimal $\G \given \TO$ for the optimal $\TO$. The first approach of
this kind was the K2 algorithm \cite{k2}, which assumed $\TO$ to be known
\emph{a priori}; other algorithms such as \cite{order1} and more recently
\cite{corani} learn the variable ordering from the data. Among these algorithms,
we will focus on the simulated annealing \cite{catnet} modification of the
Metropolis-Hastings topological ordering search covered in \cite{koller}. The
algorithm is illustrated in Algorithm \ref{algo:sann}: step \ref{step:order}
maximises $\Prob(\TO \given \D)$, while step \ref{step:conditional} maximises
$\Prob(\G \given \TO, \D)$. Hence Algorithm \ref{algo:sann} maximises
\begin{equation*}
  \Prob(\G \given \D) = \Prob(\G, \TO \given \D) = \Prob(\TO \given \D) \Prob(\G \given \TO, \D)
\end{equation*}
since the topological ordering $\TO$ is a function of $\G$. Step
\ref{step:order} generates a new topological ordering $\TO_i$ at each
iteration, which then is carried forward to the next iteration with a transition
probability $\Prob(\TO_i \given \TO_{i -1}, \beta)$ that depends on the
relative goodness-of-fit of of $\TO_i$ and $\TO_{i -1}$. The latter can
be calculated either by averaging over all possible DAGs compatible with each
topological ordering
\begin{equation}
  \Prob(\TO \given \D) \propto \Prob(\D \given \TO) \propto
    \int \Prob(\D \given \G) \Prob(\G \given \TO) \dd\G;
\end{equation}
or by finding the DAG with the best score for each topological ordering subject
to some constraints such as the maximum number of parents for each node. The
role of $\beta$ is to control the annealing schedule by gradually reducing the
transition probability.

Finally, hybrid algorithms combine the previous two approaches. They consist of
two steps, called \textit{restrict} and \textit{maximise}. In the first step, a
candidate set $\mathbf{C}_{X_i}$ of parents is selected for each node $X_i$
from $\X \setminus X_i$ using conditional independence tests. Assuming that all
$\mathbf{C}_{X_i}$ are small compared to $\X$, we are left with a smaller and
more regular space in which to search for our network structure. The second step
seeks the DAG that maximises a given network score function subject to the
constraint that the parents of each $X_i$ must be in the corresponding
$\mathbf{C}_{X_i}$.  In practice, the first step is implemented using the part
of some constraint-based algorithm that identified the skeleton of the network,
corresponding to steps \ref{step:pc1} and \ref{step:pc2} in Algorithm
\ref{algo:pc}. The second step, on the other hand, is implemented using a
score-based algorithm such as Algorithms \ref{algo:greedy} and \ref{algo:sann}
above. The best-known member of this family is the \emph{Max-Min Hill Climbing}
algorithm (MMHC) by \cite{mmhc}; two other examples are RSMAX2 from our previous
work \cite{magic14} and \HHPC{} \cite{h2pc}.

\subsection{Statistical Criteria: Conditional Independence Tests and Network Scores}
\label{sec:intro-criterion}

The choice of which statistical criterion to use in structure learning, be that
a conditional independence test or a network score, depends mainly on the
distribution of $\X$; and is orthogonal to the choice of algorithm. Here we
provide a brief overview of those in widespread use in the literature, while
referring the reader to \cite{koller} for a more comprehensive treatment.

For discrete BNs, conditional independence tests are functions of the observed
frequencies $\{n_{ijk}; i = 1, \ldots, R,$ $j = 1, \ldots, C; k = 1, \ldots,
L\}$ for any pair of variables ($X$, $Y$) given the configurations of some
conditioning variables $\Z$. In other words, $X$, $Y$ and $\Z$ take one of $R$,
$C$ and $L$ possible values for each observation. The two most common tests are
the log-likelihood ratio $\GSQ$ test and Pearson's $\XT$ test. $\GSQ$ is defined
as
\begin{equation}
  \GSQ(X, Y \given \Z)
    = 2 \log\frac{\Prob(X \given Y, \Z)}{\log\Prob(X \given \Z)}
    = 2 \sum_{i=1}^R \sum_{j=1}^C \sum_{k=1}^L
        n_{ijk} \log\frac{n_{ijk} n_{++k}}{n_{i+k} n_{+jk}},
\label{eq:g2d}
\end{equation}
where $n_{i+k} = \sum_{j = 1}^C n_{ijk}$, $n_{+jk} = \sum_{i = 1}^R n_{ijk}$ and
$n_{++k} = \sum_{i = 1}^R \sum_{j = 1}^C n_{ijk}$ are the marginal counts for
$i, k$ (summed over $i$); $j, k$ (summed over $i$); and $k$ (summed over $i$
and $j$). $\XT$ is defined as
\begin{align*}
  &\XT(X, Y \given \Z) = \sum_{i=1}^R \sum_{j=1}^C \sum_{k=1}^L
     \frac{\left(n_{ijk} - m_{ijk}\right)^2}{m_{ijk}},&
  &\text{where}&
  &m_{ijk} = \frac{n_{i+k}n_{+jk}}{n_{++k}}.
\end{align*}
Both are asymptotically equivalent\footnote{$\XT - \GSQ$ converges to zero in
probability, meaning $\Prob(|\XT - \GSQ| < \varepsilon) \to 1$ as $n \to \infty$
for any $\varepsilon > 0$ \citep{agresti}.} and have the same $\chi^2_{(R - 1)(C - 1)L}$
null distribution. Notably, $\GSQ$ is also numerically equivalent to mutual
information (they differ by a $2n$ factor).

For GBNs, conditional independence tests are functions of the partial
correlation coefficients $\rXYZ$. The log-likelihood ratio (and Gaussian mutual
information) test takes form
\begin{equation}
  \GSQ(X, Y \given \Z) = n\log(1 - \rXYZ^2) \sim \chi^2_1.
\label{eq:g2g}
\end{equation}
Other common options are the exact Student's $t$ test
\begin{equation*}
  \mathrm{t}(X, Y \given \Z) =
    \rXYZ \sqrt{\frac{n - |\Z| - 2}{1 - \rXYZ^2}}
    \sim t_{n - |\Z| - 2};
\end{equation*}
and the asymptotic Fisher's $Z$ test, defined as
\begin{equation*}
  \mathrm{Z}(X, Y \given \Z) = \log\left(\frac{1 + \rXYZ}
    {1 - \rXYZ}\right) \frac{\sqrt{n - |\Z| - 3}}{2}
  \sim N(0, 1).
\end{equation*}

As for network scores, the Bayesian Information criterion
\begin{equation}
  \BIC(\G; \D) =
    \sum_{i=1}^N \left[\; \log \Prob(\XPi) - \frac{|\T|}{2}\log n \;\right],
\label{eq:bic}
\end{equation}
is a common choice for both discrete BNs and GBNs, because it provides a simple
approximation to $\log\Prob(\G \given \D)$ that does not depend on any
hyperparameter. $\log\Prob(\G \given \D)$ is also available in closed form for
both discrete BNs and GBNs.

In discrete BNs, $\Prob(\D \given \G)$ is called the \emph{Bayesian Dirichlet}
(BD) score \cite{heckerman} and it is constructed using a conjugate Dirichlet
prior with imaginary sample size $\boldsymbol{\alpha}$ (the size of an imaginary
sample supporting the prior distribution, giving the weight given to the prior
compared to the data). It takes the form
\begin{equation}
\label{eq:bd}
  \BD(\G, \D; \boldsymbol{\alpha}) =
  \prod_{i=1}^N \BD(\XPi; \alpha_i) =
  \prod_{i=1}^N \prod_{j = 1}^{q_i}
    \left[
      \frac{\Gamma(\alpha_{ij})}{\Gamma(\alpha_{ij} + n_{ij})}
      \prod_{k=1}^{r_i} \frac{\Gamma(\aijk + n_{ijk})}{\Gamma(\aijk)}
    \right]
\end{equation}
where
\begin{itemize}
  \item $r_i$ is the number of states of $X_i$;
  \item $q_i$ is the number of configurations of $\PXi$;
  \item $n_{ij} = \sum_k n_{ijk}$, the marginal count for the $k$th parents
    configuration;
  \item the $\aijk$ are the hyperparameters of the Dirichlet distribution, and
    $\alpha_{ij} = \sum_k \aijk$, $\alpha_i = \sum_j \alpha_{ij}$.
\end{itemize}
The most common choice for the hyperparameters is $\aijk = \alpha_i/(r_i q_i)$,
which gives the \emph{Bayesian Dirichlet equivalent uniform} (BDeu) score, the
only BD score that satisfies score equivalence. It is typically used with small
imaginary sample sizes such as $\alpha_i = 1$ as suggested by \cite{koller} and
\cite{ueno}. Alternative BD scores have been proposed in \cite{suzuki16} and
\cite{pgm16,behaviormetrika18}.

As for GBNs, $\log\Prob(\D \given \G)$ is called the \emph{Bayesian Gaussian
equivalent} (BGe) score and it is constructed using a conjugate normal-Wishart
prior for $\X$ with expected values $\nnu$ (for the mean) and $T$ (for the
covariance). It takes the form \cite{kuipers}
\begin{align}
  \BG(\G, \D; \aw, \am, T, \nnu) \nonumber
    &= \prod_{i=1}^N \BG(\XPi; \aw, \am, T, \nnu) \\ \nonumber
    &= \prod_{i=1}^N \prod_{j = 1}^{q_i}
      \left( \frac{\am}{n + \am} \right)
      \frac{\Gamma\left(\frac{n + \aw - N + |\PXi| + 1}{2}\right)}
           {\pi^{n/2} \Gamma\left(\frac{\aw - N + |\PXi| + 1}{2}\right)} \\
    & \qquad\cdot
      \frac{\left|T_{X_i, \PXi}\right|^{\frac{\aw - N - |\PXi| - 1}{2}}}
           {\left|T_{\PXi}\right|^{\frac{\aw - N - |\PXi|}{2}}}
      \frac{\left|R_{\PXi}\right|^{\frac{n + \aw - N - |\PXi|}{2}}}
           {\left|R_{X_i, \PXi}\right|^{\frac{n + \aw - N - |\PXi| - 1}{2}}}
\label{eq:bge}
\end{align}
where:
\begin{itemize}
  \item $\am$ and $\aw$ are the imaginary sample sizes that give the weight of
    the normal and Wishart components of the prior compared to the sample;
  \item $R$ is the posterior covariance matrix and is given by
    \begin{align*}
      &R = T + S_n + \frac{n\am}{n + \am}
         (\xm - \nnu)^T (\xm - \nnu),&
      &\bar{\mathbf{x}} = \frac{1}{n} \sum_{i = 1}^n \mathbf{x}_i, \quad
         S_n = \sum_{i = 1}^n (\mathbf{x}_i - \xm) (\mathbf{x}_i - \xm)^T,
    \end{align*}
    where $\mathbf{x}_i$ is a complete instantiation of $\X$;
  \item $T_{X_i, \PXi}$ and $R_{X_i, \PXi}$ are the submatrices of $T$ and $R$
    corresponding to the ($X_i$, $\PXi$);
  \item similarly, $T_{\PXi}$ and $R_{\PXi}$ are the submatrices of $T$ and $R$
    corresponding to the $\PXi$.
\end{itemize}
\cite{kuipers} suggests using the smallest valid values for both imaginary
sample sizes ($\am = 1$, $\aw = N + 2$), a diagonal $T = t I_N$ with
\begin{equation*}
  t = \frac{\am (\aw - N - 1)}{\am + 1},
\end{equation*}
and $\nnu = \xm$ as a set of default values with wide applicability for the
hyperparameters.

\section{Performance as a Combination of Statistical Criteria and Algorithms}

As it may be apparent from Sections \ref{sec:intro-algorithms} and
\ref{sec:intro-criterion}, we take the view that the algorithms and the
statistical criteria they use are separate and complementary in determining the
overall behaviour of structure learning. Cowell \cite{cowell} followed the same
reasoning when showing that constraint-based and score-based algorithms can
select identical discrete BNs. He noticed that the $\GSQ$ test in \mref{eq:g2d}
has the same expression as a score-based network comparison based on the
log-likelihoods $\log\Prob(X \given Y, \Z) - \log\Prob(X \given \Z)$ if we take
$\Z = \Pi_{X}$. He then showed that these two classes of algorithms are
equivalent if we assume a fixed, known topological ordering\footnote{This
assumption is required because $\GSQ$ can only be used to test arc addition or
removal; given a fixed topological ordering these are the only two possible
single-arc operations because arc reversing any arc would change the topological
ordering of the nodes. Cowell briefly suggests in the Conclusions of
\cite{cowell} that it might be possible to relax it if it were possible to test
arc reversal in a single statistical test, as opposed to performing two separate
tests for removing an arc and adding it back in the opposite direction. However,
to the best of our knowledge no such test has been proposed so far in the
literature.} and we use log-likelihood and $\GSQ$ as matching statistical
criteria.

In this paper we will complement that investigation by addressing the following
questions:
\begin{enumerate}
  \item[\textbf{Q1}] \emph{Which of constraint-based and score-based algorithms
    provide the most accurate structural reconstruction, after accounting for
    the effect of the choice of statistical criteria?}
  \item[\textbf{Q2}] \emph{Are constraint-based algorithms faster than
    score-based algorithms, or vice-versa?}
  \item[\textbf{Q3}] \emph{Are hybrid algorithms more accurate than
    constraint-based or score-based algorithms?}
  \item[\textbf{Q4}] \emph{Are hybrid algorithms faster than
    constraint-based or score-based algorithms?}
  \item[\textbf{Q5}] \emph{Do the different classes of algorithms present any
    systematic difference in either speed or accuracy when learning small
    networks and large networks?}
\end{enumerate}
More precisely, we will drop the assumption that the topological ordering is
known and we will compare the performance of different classes of algorithms
outside of their equivalence conditions for both discrete BNs and GBNs. We
choose questions \textbf{Q1}, \textbf{Q2}, \textbf{Q3}, \textbf{Q4} and
\textbf{Q5} because they are most common among practitioners
\cite[\emph{e.g.}][]{cugnata} and researchers
\cite[\emph{e.g.}][]{mmhc,koller,spirtes2}. Overall, there is a general view in
these references and in the literature that score-based algorithms are less
sensitive to individual errors of the statistical criteria, and thus more
accurate, because they can reverse earlier decisions; and that hybrid algorithms
are faster and more accurate than both score-based and constraint-based
algorithms. These differences have been found to be more pronounced at small
sample sizes. Furthermore, score-based algorithms have been found to scale less
well to high-dimensional data.

We find two important limitations in such investigations. The first is that they
focus almost exclusively on discrete BNs, ignoring that GBNs are based on very
different distributional assumptions and thus that their conclusions will not
necessarily hold for the latter. The second is the confounding between the
choice of the algorithms and that of the statistical criteria, which makes it
impossible to assess the merits inherently attributable to the algorithms
themselves. Therefore, similarly to \cite{cowell}, we construct matching
statistical criteria in the form of pairs of scores and independence tests that
make algorithms directly comparable. Without loss of generality, consider two
DAGs $\Gp$ and $\Gm$ which differ by a single arc $X_j \rightarrow X_i$. In a
score-based approach, we can compare them using BIC from \mref{eq:bic} and
select $\Gp$ over $\Gm$ if

\begin{equation}
  \BIC(\Gp; \D) > \BIC(\Gm; D) \Rightarrow
    2\log\frac{\Prob(\XPi \cup \{X_j\})}{\Prob(\XPi)} > (|\T^{\Gp}| - |\T^{\Gm}|)\log n
\label{eq:g2bic}
\end{equation}
which is equivalent to testing the conditional independence of $X_i$ and $X_j$
given $\PXi$ using the $\GSQ$ test from \mref{eq:g2d} or \mref{eq:g2g}, just
with a different significance threshold than a $\chi^2_{1 - \alpha}$ quantile at
a pre-determined significance level $\alpha$. We will call this test $\GB$ and
use it as the matching statistical criterion for $\BIC$ to compare different
learning algorithms. In addition, we will construct a second test along the
same lines using graph posterior probabilities in order to confirm our
conclusions with a second set of matching criteria. Following \mref{eq:g2bic},
we write
\begin{equation*}
  \log\Prob(\Gp \given \D) > \log\Prob(\Gm \given \D) \Rightarrow
    \log\BF = \log\frac{\Prob(\Gp \given \D)}{\Prob(\Gm \given \D)} > 0
\end{equation*}
which decides between $\Gp$ and $\Gm$ using a Bayes factor with a threshold of
$1$, similarly to what was previously done in \cite{natori}. The resulting
($\BIC$, $\GB$) and ($\log\Prob(\G \given \D)$, $\log\BF$) will be used to
investigate discrete BNs and GBNs in the following section. An extension of
($\BIC$, $\GB$) to the family of matching criteria $(\BICg, \GSQ_{\BICg})$ will
be used to investigate GBNs learned from real-world complex data in Section
\ref{sec:climate}.

\section{Simulation Study}
\label{sec:empirical}

We address \textbf{Q1}, \textbf{Q2}, \textbf{Q3}, \textbf{Q4} and \textbf{Q5}
with a simulation study based on reference BNs from the Bayesian network
repository \cite{bnrepo}; we will later confirm our conclusions using real-world
complex climate data in Section \ref{sec:climate}. Both will be implemented
using the \emph{bnlearn} \cite{jss09} and \emph{catnet} \cite{catnet} R packages
and TETRAD \cite{tetrad} via the \emph{r-causal} R package \cite{r-causal}.

We assess the structure learning algorithms listed in Table
\ref{tab:algorithms}: three constraint-based (PC-Stable, GS, Inter-IAMB), three
score-based (tabu search, simulated annealing for BIC, GES for $\log\Prob(\G
\given \D)$) and three hybrid algorithms (MMHC, RSMAX2, \HHPC{}). For this
purpose we use the 10 discrete BNs and 4 GBNs in Table \ref{tab:networks}. For
each BN:
\begin{enumerate}
  \item We generate $20$ samples of size $n/|\Theta| = 0.1$, $0.2$, $0.5$,
    $1.0$, $2.0$, and $5.0$ to allow for meaningful comparisons between BNs
    of very different size and complexity. Intuitively, an absolute sample of
    size of, say, $n = 1000$ may be large enough to learn reliably a small
    BN with few parameters, say $|\Theta| = 100$, but it may be too small
    for a larger or denser network with $|\Theta| = 10000$. Using the relative
    sample size $n/|\Theta|$ ensures small and large sample regimes are
    consistent for different BNs.
  \item We learn $\G$ using (BIC, $\GB$) and ($\log\Prob(\G \given \D)$,
    $\log\BF$). For the latter we use the BDeu and BGe scores in \mref{eq:bd}
    and \mref{eq:bge} with the hyperparameter values suggested in Section
    \ref{sec:intro-criterion}. In addition we set a prior probability of
    inclusion of $1/(N-1)$ for each parent of each node, which is the default in
    TETRAD.
  \item We measure the accuracy of the learned DAGs using the Structural Hamming
    Distance \cite[SHD;][]{mmhc} from the reference BN scaled by the number of
    arcs $|A|$ of that BN (lower is better). This again motivated by the need to
    compare networks of different sizes: if both the reference BN and the
    learned network are sparse then we expect SHD to be $O(|A|)$, since both
    will have $O(|A|)$ arcs.
  \item We measure the speed of the learning algorithms with the number of calls
    to the statistical criterion (lower is better). This is a classic measure of
    computational complexity in BN structure learning.
\end{enumerate}

\begin{table}[hp!]
\begin{center}
  \small
  \begin{tabular}{|r|l|c|c|c|c|}
  \hline
  algorithm                      & class & discrete BNs & GBNs  & (BIC, $\GB$) & ($\log\Prob(\G \given \D)$, $\log\BF$) \\
  \hline
  PC-Stable                      & constraint-based & $\ck$        & $\ck$ & $\ck$           & $\ck$ \\
  Grow-Shrink (GS)               & constraint-based & $\ck$        & $\ck$ & $\ck$           & $\ck$ \\
  Inter-IAMB                     & constraint-based & $\ck$        & $\ck$ & $\ck$           & $\ck$ \\
  tabu search                    & score-based      & $\ck$        & $\ck$ & $\ck$           & $\ck$ \\
  simulated annealing            & score-based      & $\ck$        & $\ck$ & $\ck$           & $\times$  \\
  Greedy Equivalent Search (GES) & score-based      & $\ck$        & $\times$ & $\times$     & (only discrete BNs) \\
  Max-Min Hill Climbing (MMHC)   & hybrid           & $\ck$        & $\ck$ & $\ck$           & $\ck$ \\
  RSMAX2                         & hybrid           & $\ck$        & $\ck$ & $\ck$           & $\ck$ \\
  \HHPC                          & hybrid           & $\ck$        & $\ck$ & $\ck$           & $\ck$ \\
  \hline
  \end{tabular}
  \caption{Structure learning algorithms compared in this paper, with their
    availability in the different simulation settings.}
  \label{tab:algorithms}

  \vspace{0.5\baselineskip}

  \begin{tabular}{|l|r|r|r|l|r|r|r|}
    \hline
    discrete BN & $N$   & $|A|$ & $|\Theta|$ & discrete BN & $N$   & $|A|$ & $|\Theta|$ \\
    \hline
    ALARM       & $37$  & $46$  & $509$      & MUNIN1      & $186$ & $273$ & $15622$ \\
    ANDES       & $223$ & $338$ & $1157$     & PATHFINDER  & $135$ & $200$ & $77155$ \\
    CHILD       & $20$  & $25$  & $230$      & PIGS        & $442$ & $592$ & $5618$  \\
    HAILFINDER  & $56$  & $66$  & $2656$     & WATER       & $32$  & $66$  & $10083$ \\
    HEPAR2      & $70$  & $123$ & $1453$     & WIN95PTS    & $76$  & $112$ & $574$   \\
    \hline
  \end{tabular}

  \vspace{0.5\baselineskip}

  \begin{tabular}{|l|r|r|r|}
    \hline
    GBN & $N$   & $|A|$ & $|\Theta|$   \\
    \hline
    ARTH150    & $107$ & $150$ & $364$ \\
    ECOLI70    & $46$  & $70$  & $162$ \\
    MAGIC-IRRI & $64$  & $102$ & $230$ \\
    MAGIC-NIAB & $44$  & $66$  & $154$ \\
    \hline
  \end{tabular}
  \caption{Reference BNs from the Bayesian network repository \cite{bnrepo}
    with the respective numbers of nodes ($N$), arcs ($|A|$) and parameters
    ($|\Theta|$).}
  \label{tab:networks}

\end{center}
\end{table}

\subsection{Discrete BNs}

\begin{figure}[!p]
\begin{center}
  \includegraphics[width=\textwidth]{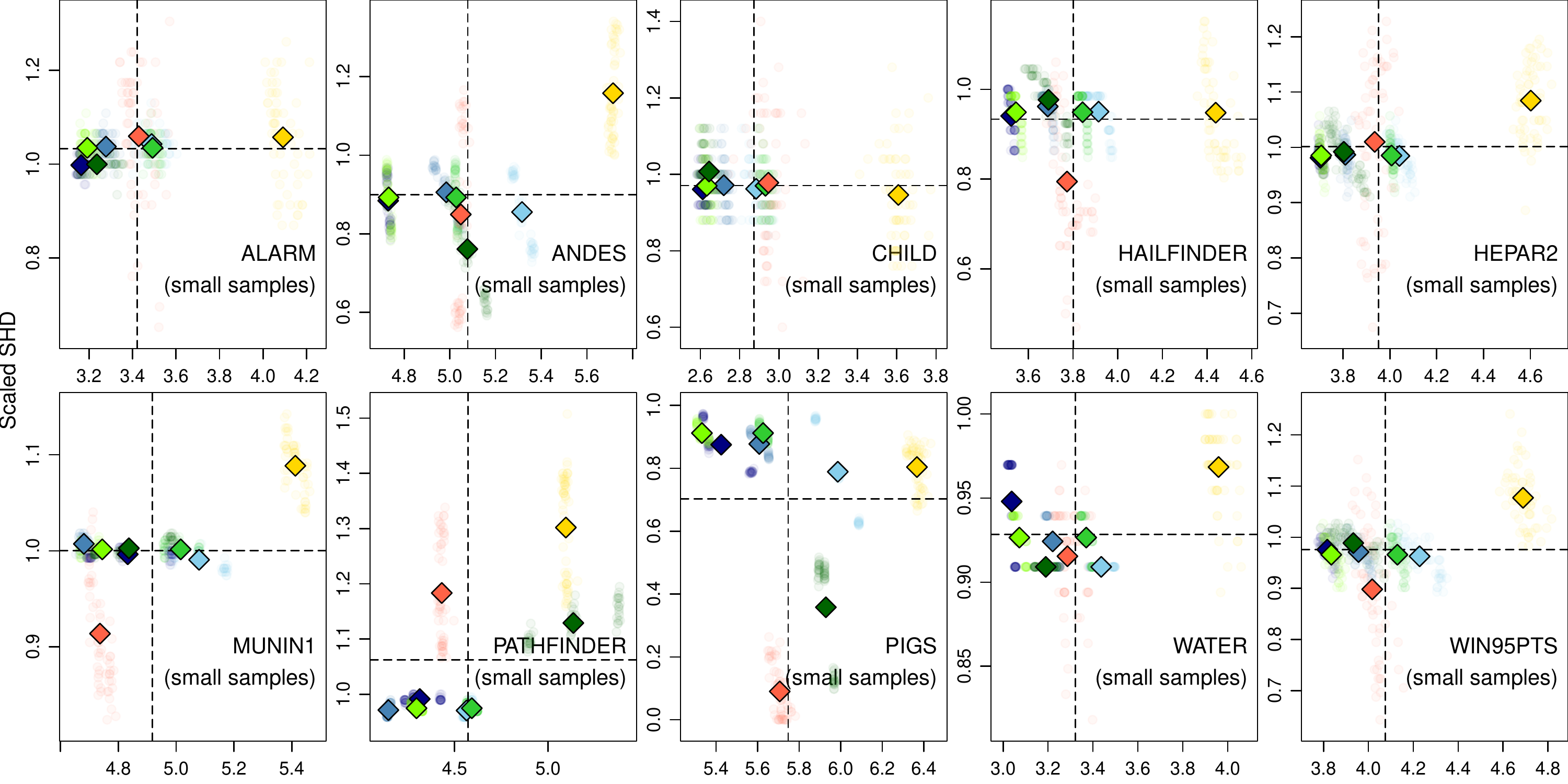}
  \includegraphics[width=\textwidth]{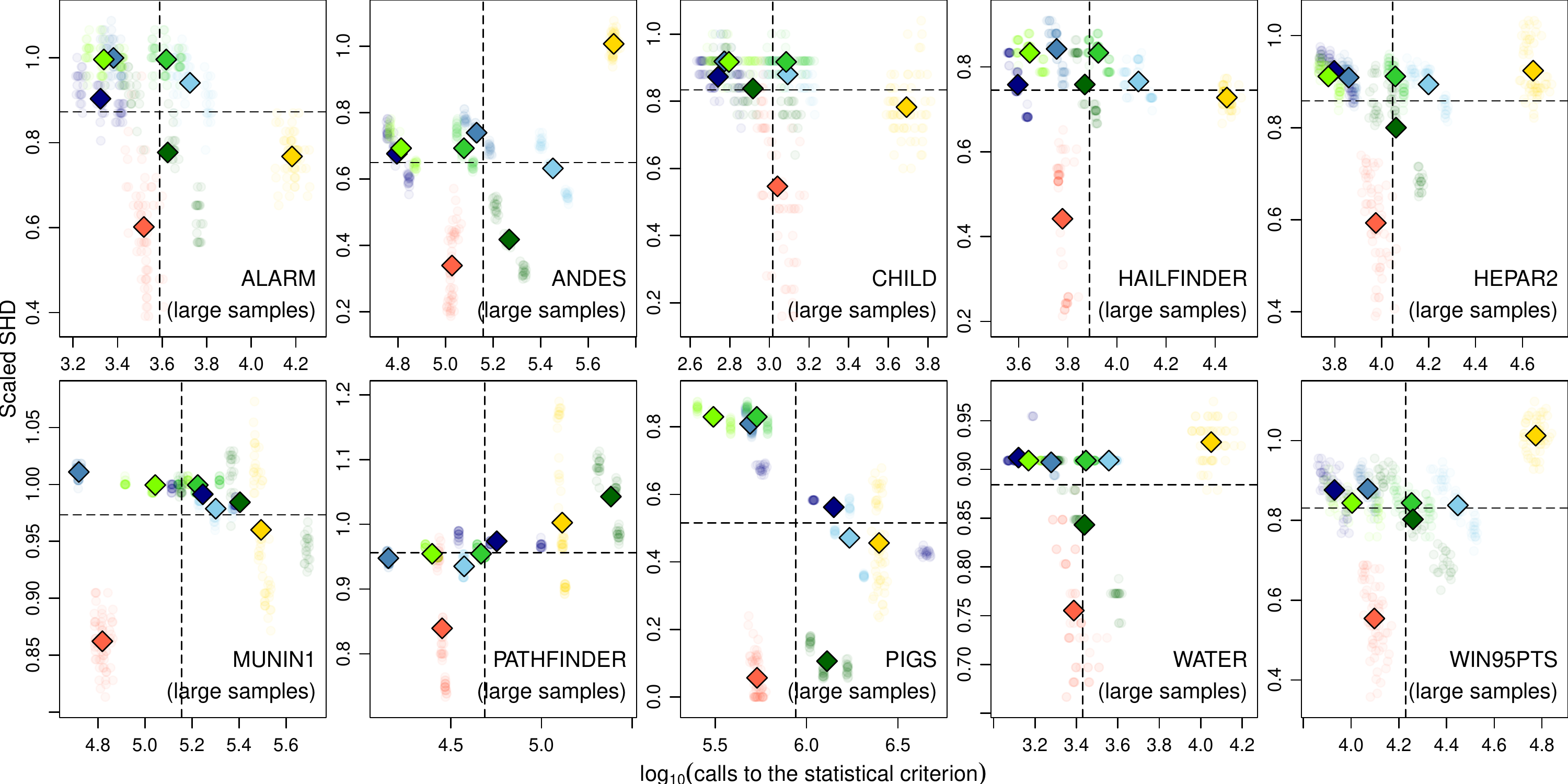}
  \caption{Scaled SHD versus speed for GS (blue), Inter-IAMB (sky blue),
    PC-Stable (navy blue), MMHC (green), RSMAX2 (lime green), \HHPC{} (dark
    green), tabu search (red) and simulated annealing (gold) and (BIC, $\GB$)
    for the discrete BNs. Shaded points correspond to individual simulations,
    while diamonds are algorithm averages.}
  \label{fig:shd-bicd}
\end{center}
\end{figure}

\begin{figure}[!p]
\begin{center}
  \includegraphics[width=\textwidth]{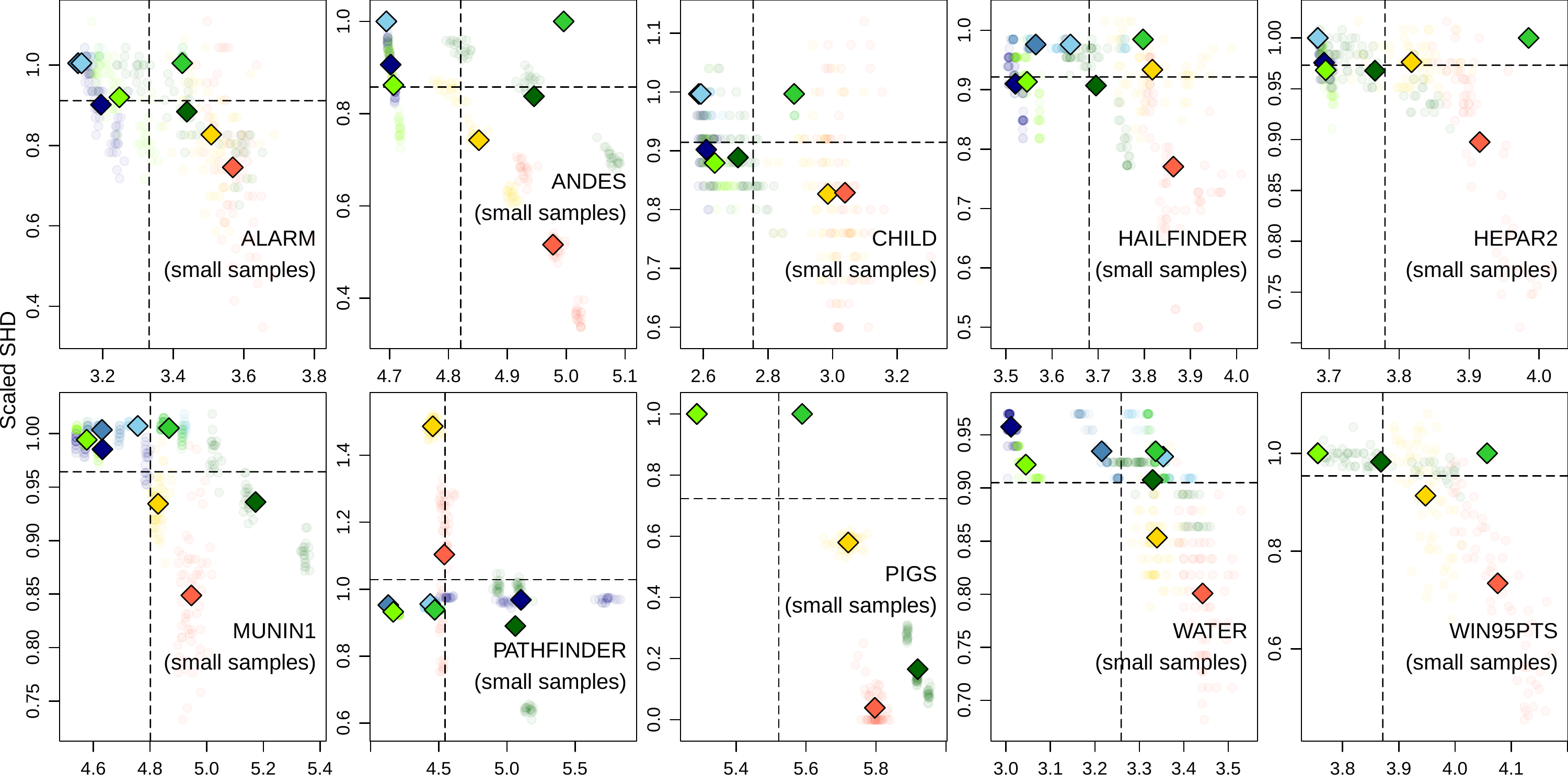}
  \includegraphics[width=\textwidth]{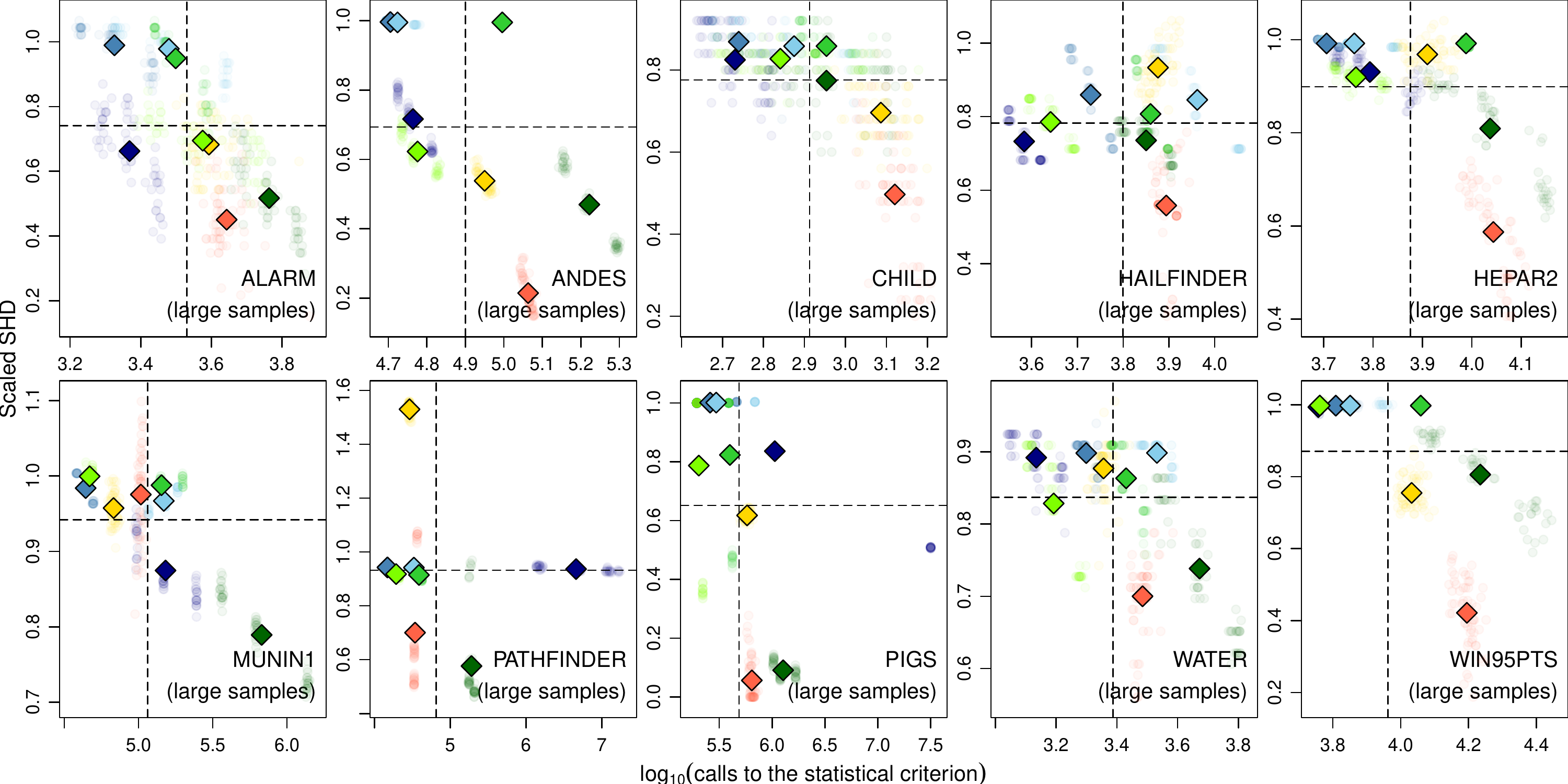}
  \caption{Scaled SHD versus speed for GS (blue), Inter-IAMB (sky blue),
    PC-Stable (navy blue), MMHC (green), RSMAX2 (lime green), \HHPC{} (dark
    green), tabu search (red) and GES (gold) and ($\log\Prob(\G \given \D)$,
    $\log\BF$) for the discrete BNs. Shaded points correspond to individual
    simulations, while diamonds are algorithm averages.}
  \label{fig:shd-bde}
\end{center}
\end{figure}

The results for discrete networks are illustrated in Figure \ref{fig:shd-bicd}
for (BIC, $\GB$) and in Figure \ref{fig:shd-bde} for ($\log\Prob(\G \given \D)$,
$\log\BF$). Results for small samples ($n/|\Theta| < 1$) and large samples
($n/|\Theta| \geqslant 1$) are shown separately in each figure. For ease of
interpretation, we divide each panel in four quadrants corresponding to ``fast,
inaccurate'' (top left), ``slow, inaccurate'' (top right), ``slow, accurate''
(bottom right) and ``fast, accurate'' (bottom, left) algorithms with respect to
the overall mean value of the scaled SHD ($y$ axis) and the number of calls to
the statistical criterion ($x$ axis, on a $\log_{10}$-scale). Algorithms are
grouped visually by colour: constraint-based algorithms are in shades of blue,
hybrid algorithms are in shades of green and score-based algorithms are in warm
colours (yellow, red).

Using (BIC, $\GB$) we find that:
\begin{itemize}
  \item Simulated annealing is the slowest algorithm for 9/10 BNs when applied
    to small samples, and for 9/10 BNs when applied to large samples; only
    \HHPC{} is slower, and only for PATHFINDER. At the same time, simulated
    annealing also has the highest scaled SHD for 7/10 BNs for small samples,
    and for 4/10 BNs for large samples. Overall, it is located in the top right
    panel (``slow, inaccurate'') in 14/20 combinations of BNs and sample sizes.
  \item On the other hand, tabu search has the lowest scaled SHD for 4/10 BNs
    for small samples and for 10/10 BNs for large samples. It is also in the
    bottom left quadrant (``fast, accurate'') in 16/20 combinations of BNs and
    sample sizes.
  \item The scaled SHD of hybrid algorithms is comparable to that of
    constraint-based algorithms for all sample sizes and BNs. For small samples
    it is approximately equal to $1$ for both classes of algorithms because they
    learn nearly empty networks; 75\% of them have less than $0.2|A|$ arcs, so
    the SHD is driven by the number of false negative arcs. For large samples,
    scaled SHD is in the $(0.8, 1)$ range, which suggests the accuracy of
    learning improves very slowly as the sample size increases.
  \item The scaled SHD of constraint-based algorithms is comparable to or better
    than that of score-based algorithms for small sample sizes in 7/10 BNs, but
    for large samples tabu search is more accurate in 10/10 BNs. This suggests
    that the accuracy of learning of tabu search improves more quickly than that
    of constraint-based algorithms; and of hybrid algorithms as well, since
    their performance is similar.
  \item While there is no consistent overall ranking of constraint-based and
    hybrid algorithms in terms of accuracy and speed, RSMAX2 and PC-Stable are
    among the fastest two in 15/20 combinations of BNs and sample sizes.
    \HHPC{}, on the other hand, has the smallest scaled SHD in 13/20 BNs.
\end{itemize}

The performance of the learning algorithms is broadly the same when replacing
($\log\Prob(\G \given \D)$, $\log\BF$) with (BIC, $\GB$). Given the lack of
suitable software, we benchmark GES instead of simulated annealing as the second
score-based algorithm under consideration. The main differences we observe are:
\begin{itemize}
  \item Tabu search has the lowest scaled SHD algorithm for 9/10 BNs in small
    samples, and in 8/10 BNs in large samples, but at the same time it is one
    of the slowest two algorithms for 15/20 combinations of BNs and sample
    sizes.
  \item GES is always faster than tabu search, but also has a higher scaled
    SHD in 18/20 combinations of BNs and sample sizes.
\end{itemize}

 \clearpage

 \clearpage

\subsection{GBNs}

\begin{figure}[!p]
\begin{center}
  \includegraphics[width=\textwidth]{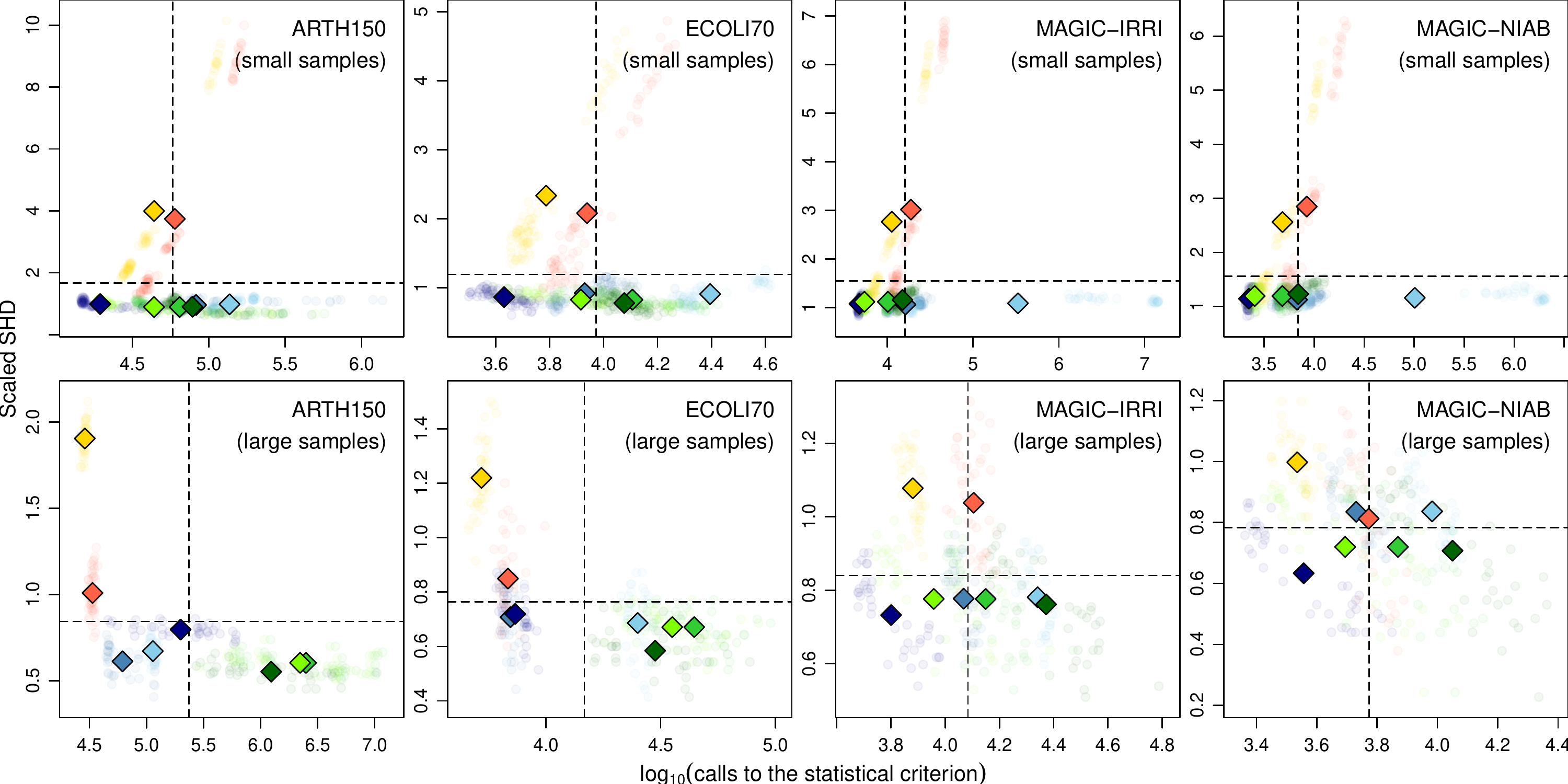}
  \caption{Scaled SHD versus speed for GS (blue), Inter-IAMB (sky blue),
    PC-Stable (navy blue), MMHC (green), RSMAX2 (lime green), \HHPC{} (dark
    green), tabu search (red) and simulated annealing (gold) and (BIC, $\GB$)
    for the GBNs.  Shaded points correspond to individual simulations, while
    diamonds are algorithm averages.}
  \label{fig:shd-bicg}
\end{center}
\begin{center}
  \includegraphics[width=\textwidth]{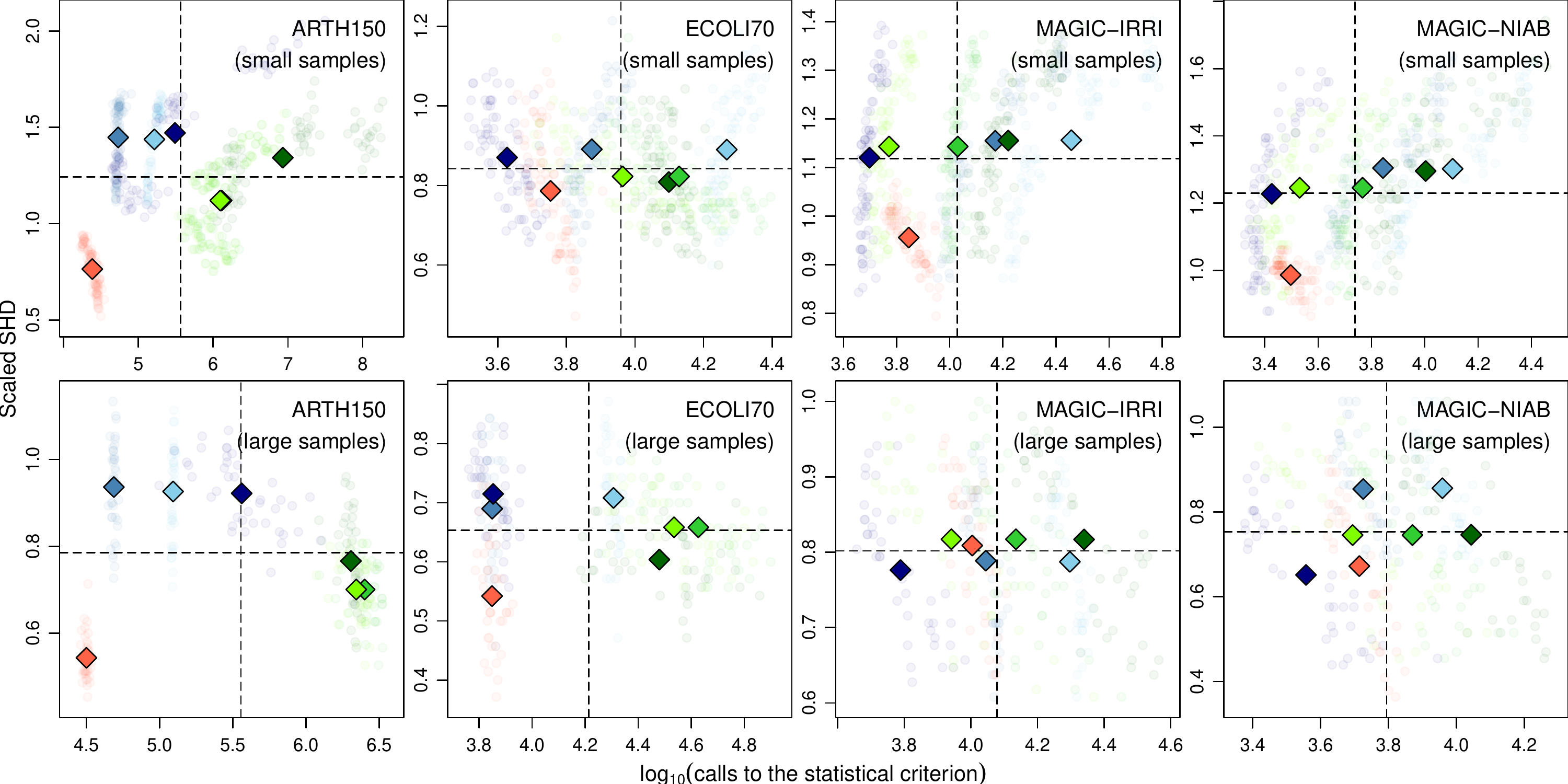}
  \caption{Scaled SHD versus speed for GS (blue), Inter-IAMB (sky blue),
    PC-Stable (navy blue), MMHC (green), RSMAX2 (lime green), \HHPC{} (dark
    green), tabu search (red) and ($\log\Prob(\G \given \D)$, $\log\BF$) for the
    GBNs.  Shaded points correspond to individual simulations, while diamonds
    are algorithm averages.}
  \label{fig:shd-bge}
\end{center}
\end{figure}

The results for GBNs are shown in Figure \ref{fig:shd-bicg} for (BIC, $\GB$),
and in Figure \ref{fig:shd-bge} for ($\log\Prob(\G \given \D)$, $\log\BF$). From
the simulations with (BIC, $\GB$), we observe that:
\begin{itemize}
  \item Tabu search and simulated annealing have a larger scaled SHD than both
    constraint-based and hybrid algorithms for all combinations of BNs and
    sample sizes. This can be attributed to the fact that the networks learned
    by tabu search and simulated annealing have a much larger number of arcs
    (between $10|A|$ and $2|A|$ for small samples, between $2|A|$ and $|A|$
    for large samples) compared to those learned by constraint-based and hybrid
    algorithms (between $0.1|A|$ and $0.8|A|$ for small samples, and between
    $0.5|A|$ and $|A|$ for large samples); many of those arcs will be false
    positives and thus increase SHD.
  \item Constraint-based and hybrid algorithms have very similar scaled SHDs
    for all combinations of BNs and sample sizes.
  \item While scaled SHD for large samples is about 40\% smaller compared to
    small samples for constraint-based and hybrid algorithms, tabu search and
    simulated annealing show a much larger improvement in accuracy (50\% to 66\%
    reduction in scaled SHD) since they start from a much worse accuracy.
  \item As was the case for discrete BNs, there is no consistent ranking of
    constraint-based and hybrid algorithms in terms of speed, PC-Stable and
    RSMAX2 are the two fastest algorithms in 7/8 combinations of BNs and sample
    sizes.
\end{itemize}

The results from the simulations performed using ($\log\Prob(\G \given \D)$,
$\log\BF$) paint a similar picture but for three important points:
\begin{itemize}
  \item Due to the lack of available software, the only score-based algorithm
    which could be used with BGe was the tabu search implementation in
    \emph{bnlearn}. This limits the conclusions that can be made from this set
    of simulations.
  \item Tabu search is in the bottom left quadrant (``fast, accurate'') in 7/8
    combinations of BNs and sample sizes, where it is also the algorithm with
    the lowest scaled SHD.
  \item While PC-Stable is still one of the two fastest among constraint-based
    and hybrid algorithms in 8/8 combinations of BNs and sample size, the same
    is true for RSMAX2 in only 4/8 combinations.
\end{itemize}

 \clearpage

\subsection{Small Networks versus Large Networks}
\label{sec:small_vs_large}

From the simulations above we can look into \textbf{Q5} as well. For this
purpose we define a ``small network'' as a BN with less than $50$ nodes, and a
``large network'' as a BN with more than $50$ nodes. Hence, the former include
ALARM, CHILD, WATER, ECOLI70 and MAGIC-NIAB; and the latter include ANDES,
HAILFINDER, HEPAR2, MUNIN1, PATHFINDER, PIGS, WIN95PTS, ARTH150 and MAGIC-IRRI.
Making this distinction based on the number of nodes is imperfect at best, since
networks of similar size can have vastly different numbers of parameters and
thus very different levels of complexity. However, it provides a categorisation
of networks that is intuitive to practitioners and that can be used when
$|\Theta|$ is unknown. In practical applications, if we assume that the discrete
BN we are trying to learn is uniformly sparse\footnote{There is no universally
accepted threshold on the number of arcs for a DAG to be called ``sparse'';
typically it is taken to have $O(cN)$ arcs, with $c$ between $1$ and $5$. A
``uniformly sparse'' DAG will have these arcs well spread among the nodes; or
equivalently, each node will have a bounded in-degree with a bound at most as
large as $c$.} and that each variable takes at most $l$ values, each local
distribution will have $O(l^{|\PXi| + 1})$ parameters and we can estimate
$|\Theta|$ with $O(Nl^{c + 1})$ taking $|\PXi| \leqslant c$ for all $X_i$.
As for GBNs, $|\Theta|$ is proportional to the number of arcs and can be
estimated as $O(cN)$; which is even more closely aligned with the number of
nodes.

Interestingly, we do not notice any systematic change in the rankings of the
learning algorithms either in terms of speed or accuracy between the two
groups of BNs. All the considerations we have made above for discrete BNs and
GBNs hold equally for small and large networks. This is important to note
because:

\begin{itemize}
  \item Different algorithms have different computational complexities, as
    measured by the expected number of calls to statistical criteria with
    respect to $N$; which may have meant that their ranking in terms of speed
    might have been different between large and small networks.
  \item Various algorithms compute different sequences of conditional
    independence tests and network scores, and thus have varying levels of
    robustness against errors in the learning process. When the matching
    statistical criteria erroneously include or exclude an arc from the network,
    different algorithms are more or less likely to erroneously include or
    exclude other arcs incident on the same nodes, which may have lead to
    important variations in the relative speed and scaled SHDs of the
    algorithms.
\end{itemize}

\section{Real-World Complex Data: A Climate Case Study}
\label{sec:climate}

In this section we address \textbf{Q1}, \textbf{Q2}, \textbf{Q3}, \textbf{Q4}
and \textbf{Q5} for real-world data considering a climate case study where
dependencies of various orders coexist. Climate data has recently attracted a
great deal of interest due to the potential application of networks to analyse
the underlying complex spatial structure \citep{fountalis_spatio-temporal_2014}.
This includes spatial dependence among nearby locations (first-order), but also
long-range (higher-order) spatial dependencies connecting distant regions in the
world, known as \emph{teleconnections} \cite{tsonis_role_2008}.  These
teleconnections represent large-scale oscillation patterns---such as the El
Ni\~no Southern Oscillation (ENSO)---which modulate the synchronous behaviour of
distant regions \cite{PhysRevLett.100.228501}. The most popular climate network
models in the literature are \emph{complex networks} \cite{tsonis_what_2006},
which are easy to build since they are based on pairwise correlations (arcs are
established between pairs of stations with correlations over a given threshold)
and provide topological information in the network structure (\emph{e.g.} highly
connected regions). BNs have been proposed as an alternative methodology for
climate networks that can model both marginal and conditional dependence
structures and that allows probabilistic inference \cite{cano}. However,
learning BNs from complex data is computationally more demanding and choosing an
appropriate structure learning algorithm is crucial. Here we consider an
illustrative climate case study modelling global surface temperature. We adapt
the matching score and independence test (BIC, $\GB$) to the family of matching
scores and independence tests ($\BICg$, $\GSQ_{\BICg}$), suitable for complex
data, and we reassess the performance of the learning methods used in Section
\ref{sec:empirical}.

\subsection{Data and Methods}

\begin{figure}[p!]
  \begin{center}
    \includegraphics[width=\textwidth]{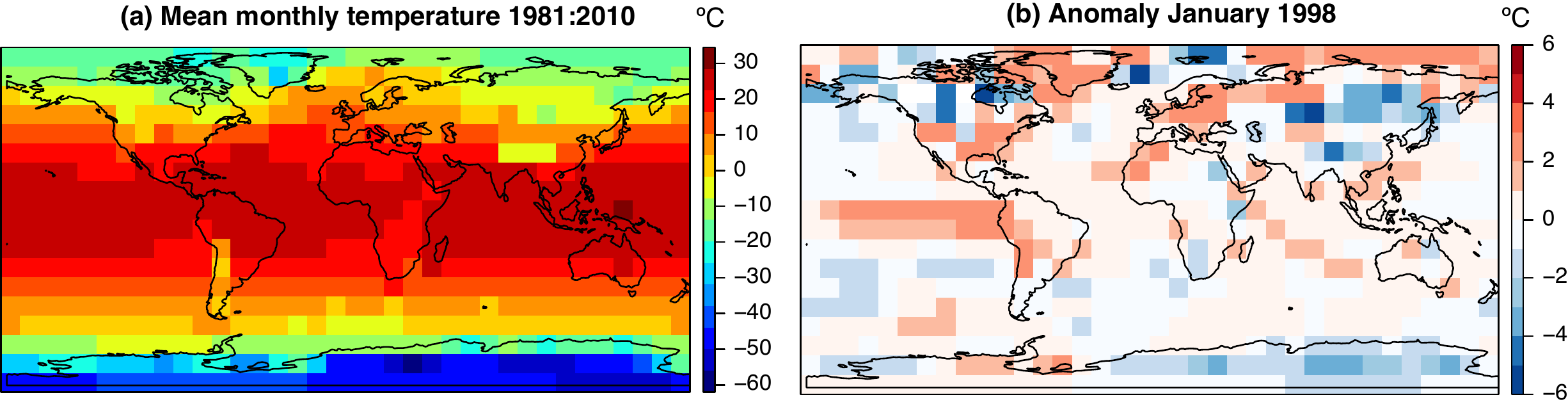}
    \caption{(a) Global mean temperature from 1981 to 2010 on a global $10^\circ$
      grid from the NCEP reanalysis. (b) Anomaly for January 1998 (strong El
      Ni\~no episode).}
    \label{fig:spatial}
  \end{center}
\end{figure}

We use monthly surface temperature values on a global $10^\circ$-resolution
(approx. 1000 km) regular grid for a representative climatic period (1981 to
2010), as provided by the NCEP/NCAR
reanalysis\footnote{\url{https://www.esrl.noaa.gov/psd/data/gridded/data.ncep.reanalysis.html}}.
Figure \ref{fig:spatial} shows the mean temperature (climatology) for the whole
period as well as the anomaly (difference from the mean 1981-2010 climatological
values) for a particular date (January 1998, representing a strong El Ni\~no
episode with high tropical Pacific temperatures).

The surface temperature at each gridpoint is assumed to be normally distributed;
hence we choose to learn GBNs in which nodes represent the (anomaly of) surface
temperature at the different gridpoints and arcs represent spatial dependencies.
Thus, we define $X_i$ as the monthly anomaly value of the temperature at
location $i$ for a period of 30 years ($n = 30 \times 12 = 360$). The anomaly
value is obtained by removing the mean annual cycle from the raw data
(\emph{i.e.} the 30-year mean monthly values) month by month. The location of a
gridpoint $i$ is defined by its latitude and longitude. Hence the node set $\X$
in the GBN is characterised as $\X = \{X_1, \dots, X_N\}$ with $N = 18 \times 36
= 648$.

In line with Section \ref{sec:empirical}, we assess two constraint-based
algorithms (PC-Stable, GS), two score-based algorithms (tabu search and hill
climbing, HC) and two hybrid algorithms (MMHC, \HHPC{}). Note, however, that in
this case the sample size is fixed to what was considered a ``small sample''
even for a DAG with no arcs: $n/|\Theta| \leqslant 360/(648 \times 2) = 0.28$.

The complex spatial dependence structure of climate data is characterised by
both local and distant (teleconnected) dependence patterns. Local dependencies
are strong since they are the result of the short-term evolution of atmospheric
thermodynamic processes. Distant teleconnected dependencies---resulting from
large-scale atmospheric oscillation patterns---are in general weaker, but they
are key for understanding regional climate variability. The various-order
dependencies in complex data are challenging for BN structure learning
algorithms and have made it necessary to introduce some adjustments in the
methodology compared to Section \ref{sec:empirical}. We show in Section
\ref{sec:constr_complex} that constraint-based algorithms are problematic when
using the $\GB$ independence test as defined in \mref{eq:g2bic}. To improve the
performance of constraint-based algorithms for complex data we introduce below
the family of extended $\BIC$ scores and independence tests. The extension makes
constraint-based, score-based and hybrid algorithms directly comparable for
complex data.

\subsubsection{Limitations of Constraint-Based Algorithms: Extended BIC for
Complex Data}
\label{sec:constr_complex}

\begin{figure}[p!]
  \begin{center}
    \includegraphics[width=\textwidth]{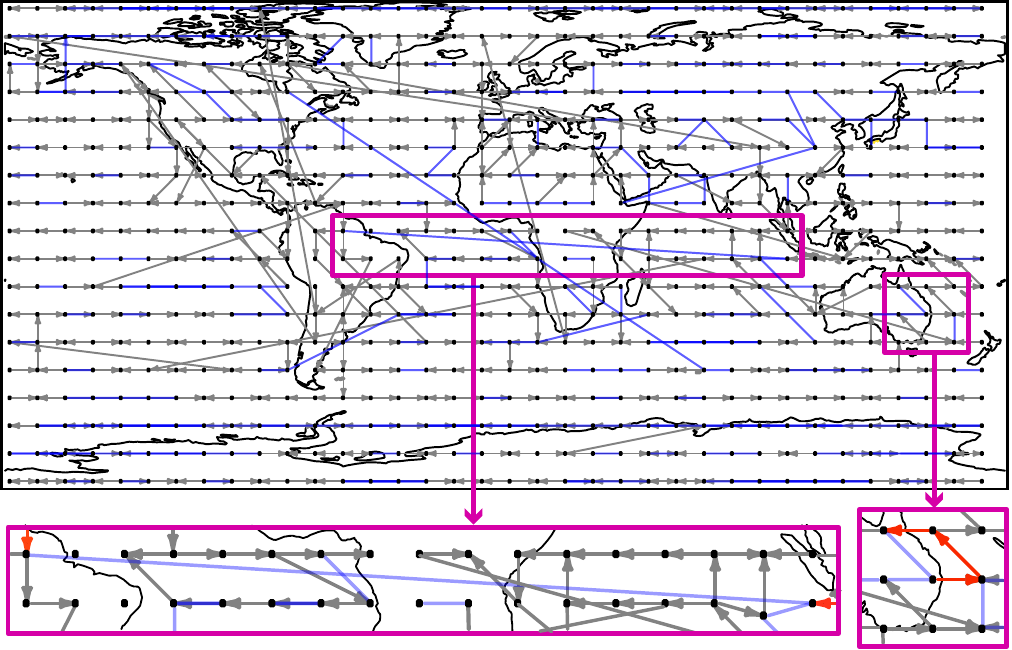}
    \caption{Partially directed graph produced by GS at the end of step
       \ref{step:pc3} with BIC. Grey arcs are directed, blue arcs are undirected
       arcs whose directions are to be set in step \ref{step:pc4}. The bottom
       left panel illustrates a case in which step \ref{step:pc4} fails to set
       arc directions between teleconnected regions. The bottom right panel
       illustrates a similar case in a highly connected region.  Short-range
       arcs in the latter and long-distance arcs in the former can not be set in
       step \ref{step:pc4} without introducing a directed cycle or a new
       v-structure (conflicting directions are shown in red).}
    \label{fig:und}
  \end{center}
\end{figure}

The heuristics that underlie constraint-based algorithms (PC-Stable and GS) and
the $\GB$ independence test, which does not enforce sparsity, are a problematic
combination when learning a CPDAG from complex data. We illustrate how and where
problems arise using climate data as an example. The algorithms first discover
highly connected local regions and some large distance arcs (Algorithm
\ref{algo:pc}, step \ref{step:pc2}). Then the algorithms attempt to identify
v-structures (step \ref{step:pc3}). This is done directly, in the case of
PC-Stable, by applying independence tests for two nodes with a common neighbour
which is not in one of their d-separating sets; and indirectly, in the case of
GS, by identifying the parents and children in the Markov blanket. In either
case, since $\GB$ does not explicitly enforce sparsity, locally connected
regions are dense (step \ref{step:pc2}) and, due to the low sample size, $\GB$
may also learn conflicting directions for the same arcs within each locally
connected region (step \ref{step:pc3}). Even though we can try to address these
conflicts with simple heuristics, such as prioritising arc directions in which
$\GB$ shows the strongest confidence, v-structures are likely to be identified
incorrectly.  Furthermore, such errors are bound to cascade in step
\ref{step:pc4} when propagating arc directions to produce a final DAG. In the
worst case, the algorithms may not be able to set the remaining arc directions
in and between highly connected regions without creating cycles or new
v-structures; an example of such a situation is shown in Figure \ref{fig:und}.
In this case the partially directed acyclic graph (PDAG) that was learned by the
algorithm in step \ref{step:pc3} does not represent an equivalence class of
DAGs, and cannot be completed into a valid CPDAG in step \ref{step:pc4}. The
learned PDAG does not encode any underlying probabilistic model and will be
referred to as an invalid CPDAG.

In order to construct an appropriate pair of matching criteria that allow
constraint-based algorithms to return valid CPDAGs for complex data, we
introduce an extended version of BIC that can produce different levels of
sparsity in the graph. The extended BIC comes with an additional regularisation
coefficient $\gamma \in \mathbb{R}^+$ that penalises the number of parameters
in the BN; which in turn are proportional to the number of arcs in the graph.
Large values of $\gamma$ thus reduce the probability of errors in steps
\ref{step:pc3} and \ref{step:pc4} for constraint-based algorithms. We refer to
this family of scores as $\BICg$, with $\BICg = \BIC$ if $\gamma = 0$, defined
as
\begin{equation*}
  \BICg(\G; \D) =
    \sum_{i=1}^N \left[ \log \Prob(\XPi) - |\T|\left(\frac{\log n}{2} - \gamma
\log N\right) \right].
\end{equation*}

We have chosen to scale $\gamma$ with the factor $|\T|\log N$ as in the EBIC
score from \cite{10.2307/24310025} due to its effectiveness in feature
selection. From $\BICg$ we then construct the corresponding independence test
$\GSQ_{\BICg}$ as follows:
\begin{equation*}
  \BICg(\Gp; \D) > \BICg(\Gm; D) \Rightarrow
  2\log\frac{\Prob(\XPi \cup \{X_j\})}{\Prob(\XPi)} > (|\T^{\Gp}| - |\T^{\Gm}|)(2\gamma \log N +\log n).
\end{equation*}

In our analysis, step \ref{step:pc4} in Algorithm \ref{algo:pc} did not produce
valid CPDAGs at all for $\gamma = 0$, and not in general for every $\gamma > 0$.
We refer to the range of $\gamma$s for which an algorithm can return valid
CPDAGs, which can then be extended into DAGs, as the \emph{parameter range} of
the algorithm. The matching statistical criteria ($\BICg$, $\GSQ_{\BICg}$)
allow us to compare the networks learned by all algorithms along their
parameter range.

Motivated by the above, we proceed as in Section \ref{sec:empirical} but with
the following changes:
\begin{enumerate}
  \item We generate 5 permutations of the order of the variables in the data to
    cancel local preferences in the learning algorithms
    \cite[see \emph{e.g.}][]{colombo}.
  \item From each permutation, we learn $\G$ using ($\BICg$, $\GSQ_{\BICg}$) for
    different values of $\gamma \in [0, 50]$.
  \item Since we do not have a ``true'' model to use as a reference, we measure
    the accuracy of learned BNs along the parameter range of the algorithm by
    their log-likelihood. We also analyse the long-distance arcs
    (teleconnections) established in the DAGs; and we assess their suitability
    for probabilistic inference by testing the conditional probabilities
    obtained when introducing some El Ni\~no-related evidence. Finally we
    analyse the conditional dependence structure by the relative amount of
    unshielded v-structures\footnote{An \emph{unshielded} v-structure is a
    pattern of arcs $X_i \rightarrow X_j \leftarrow X_k$ in which $X_i$ and
    $X_k$ are not connected by an arc. In contrast, in a \emph{shielded}
    v-structure there is a directed arc between $X_i$ and $X_k$.} in the
    network.
  \item We measure the speed of the learning algorithms with the number of calls
    to the statistical criterion.
\end{enumerate}

\begin{figure}[p!]
\begin{center}
  \includegraphics[width=1\textwidth]{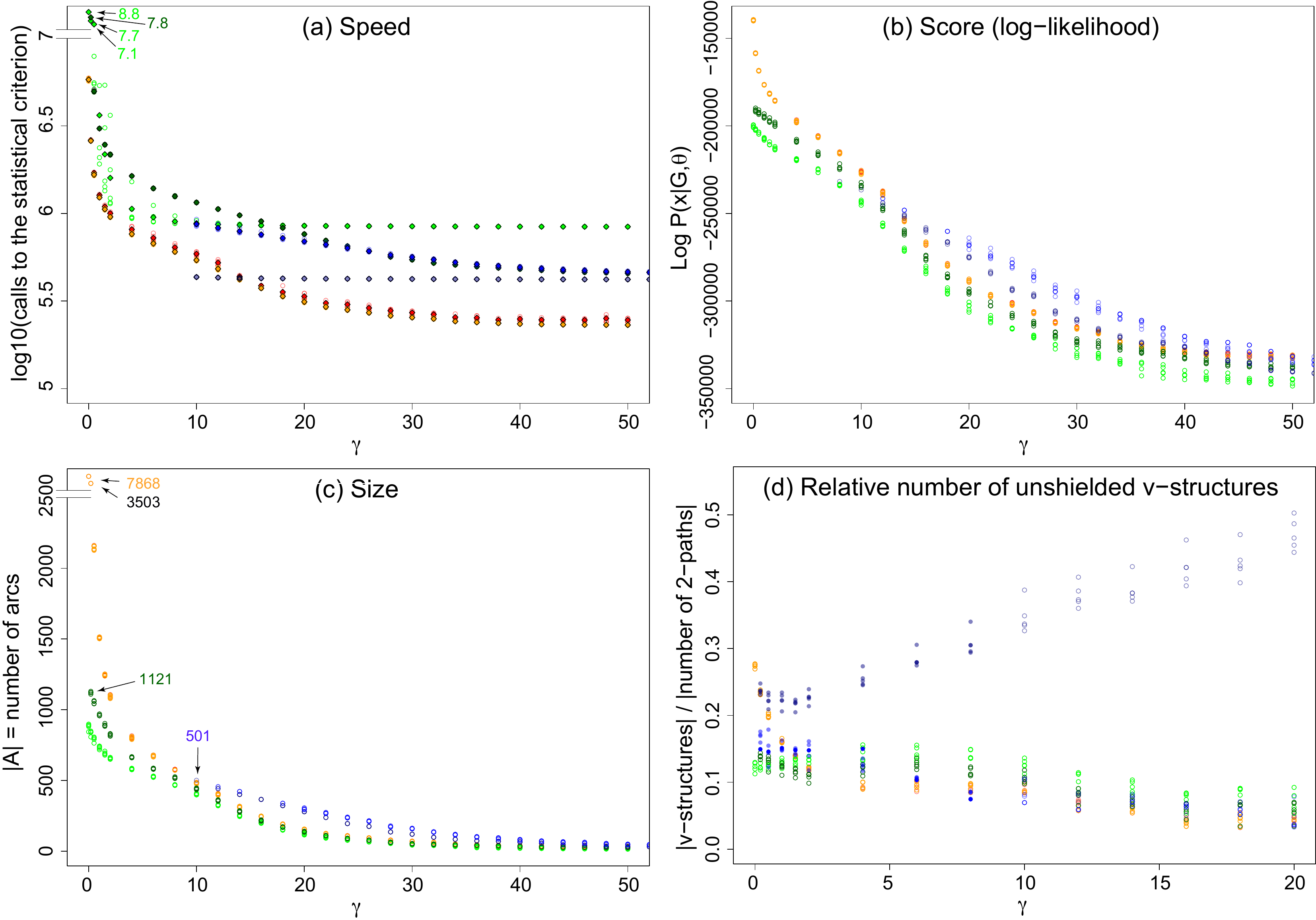}
  \caption{(a) Speed, (b) goodness of fit (log-likelihood), (c) number of arcs,
      (d) conditional dependence structure (unshielded v-structures) for
      different values of $\gamma$ learned by GS (blue), PC-Stable (navy), MMHC
      (green), \HHPC{} (dark green), tabu search (red) and HC (orange). Note
      that orange results are on top of red ones in some cases. For clarity
      panel (a) includes the mean of the 5 realisation results for each
      $\gamma$. Labelled points in (a) have means returned by MMHC and \HHPC{}
      for $\gamma \in \{0,0.2,0.5\}$ that are in speed-range higher than $7.0$.
      Labelled points in (c) represent the biggest networks of tabu for $\gamma
      \in \{0,0.2\}$ and the biggest networks found by \HHPC{} and PC-Stable (to
      be analysed in Figure \protect \ref{fig:graphs}). Filled dots in (d)
      indicate invalid equivalence classes (CPDAGs).}
  \label{fig:results}
\end{center}
\end{figure}

\subsection{Results}

\begin{figure}[!p]
\begin{center}
  \includegraphics[width=\textwidth]{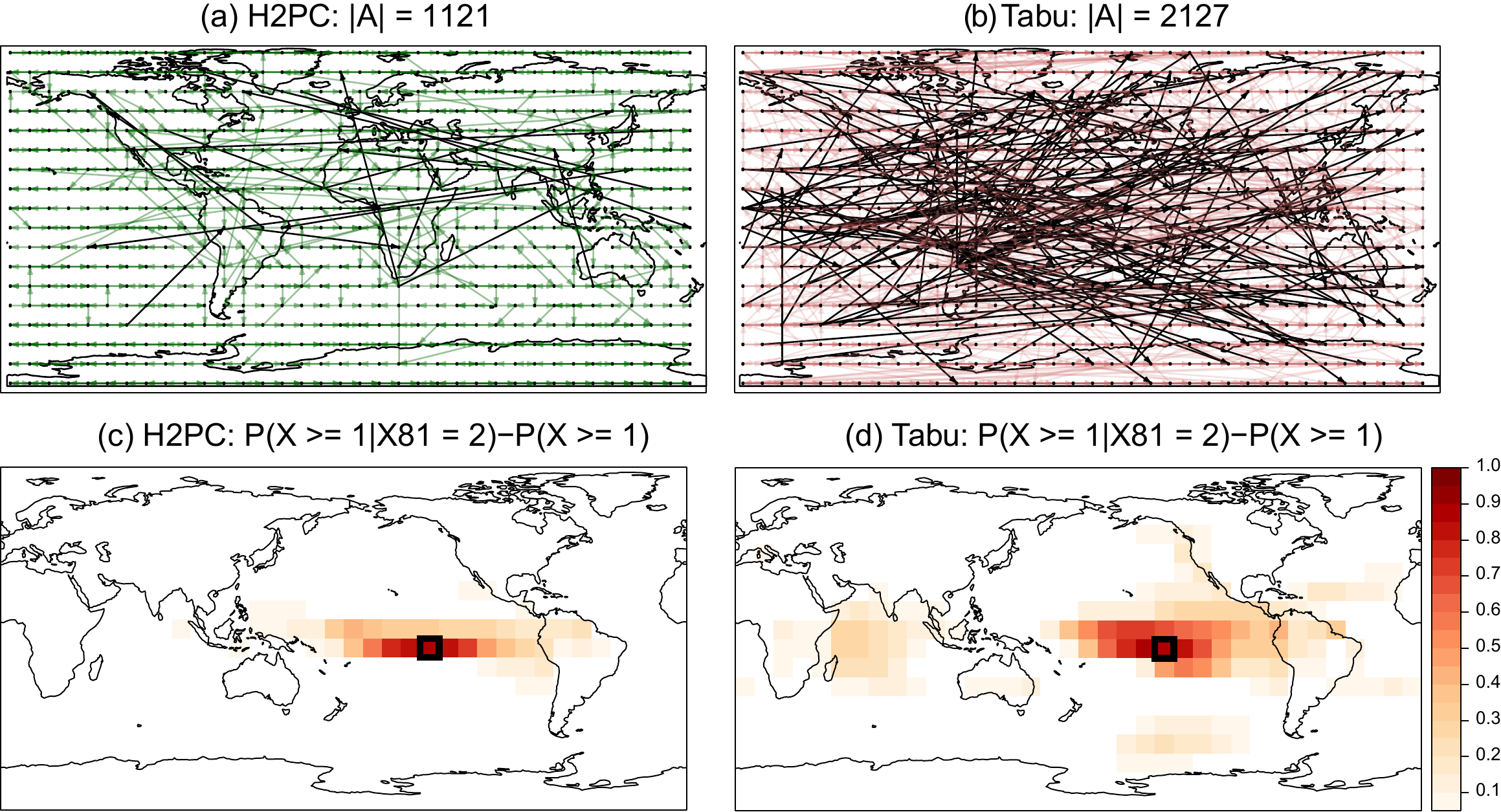}
  \caption{DAGs learned by (a) tabu search ($\gamma = 0.5$) and (b) \HHPC{}
    ($\gamma = 0.2$). Long range links (representing teleconnections) are shown
    in black. (c) and (d) show the differences of the conditional and marginal
    probabilities obtained with both Bayesian networks after propagation of
    $X_{81}=2$ (denoted with a black box), simulating El Ni\~no conditions; the
    graph obtained with tabu encodes well known teleconnection regions (e.g.
    Indian ocean) for this evidence.}
  \label{fig:graphs}
\end{center}
\end{figure}

Figure \ref{fig:results}(a-c) shows the performance (speed, goodness of fit,
number of arcs) of various structure learning algorithms as a function of
$\gamma$, using the same colours as in Figure \ref{fig:shd-bicg} (with the
exception of hill climbing, which is new in this figure and is shown in orange).
Figure \ref{fig:results}(d) shows the conditional dependence structure
(characterised by relative number of unshielded v-structures) of the CPDAGs
returned by the algorithms as a function of $\gamma$. Filled dots for PC-Stable
and GS denote invalid CPDAGs. Figure \ref{fig:results}(d) is discussed
separately at the end of this section. Figure \ref{fig:graphs} (a-b) shows the
the two representative networks from \HHPC{} and tabu search that are
highlighted with a label in Figure \ref{fig:results}(c) overlaid with the world
map. This figure also compares the suitability of the learned BNs for
probabilistic inference by propagating an El Ni\~no-like evidence ($X_{81}=2$,
\emph{i.e.} warm temperatures in the corresponding gridbox in tropical Pacific).

From the networks learned with ($\BICg$, $\GSQ_{\BICg}$) for $\gamma \in [0,50]$,
we observe that:

\begin{itemize}
  \item GS and PC-Stable produce BNs with the highest log-likelihood for large
    values of $\gamma$ ($\gamma \geq 10$, Figure \ref{fig:results}(b)).
  \item However, GS and PC-Stable do not produce valid CPDAGs for small values
    of $\gamma$ ($\gamma < 10$); and for $\gamma \geq 10$ they learn CPDAGs with
    at most $501$ arcs (smaller than the number of nodes) and no
    teleconnections, which are not useful for inference. (A constraint-based
    network is therefore excluded of Figure \ref{fig:graphs}.)
  \item \HHPC{} and MMHC exhibit the poorest log-likelihood values when
    $\gamma \geq 10$. However, in contrast with PC-Stable and GS, for $\gamma <
    10$ they do return valid CPDAGs resulting in a maximum number of $1121$ arcs
    for \HHPC{}, including some teleconnections (Figure \ref{fig:graphs}(a)).
  \item Inference on networks learned by hybrid and constraint-based algorithms
    does not highlight altered probabilities of high temperatures in the Indian
    Ocean when El Ni\~no-like evidence is given (Figure \ref{fig:graphs}(c),
    largest \HHPC{} network). High temperatures in the Indian ocean, induced by
    atmospheric teleconnection, are typical when El Ni\~no occurs as was
    illustrated in figure (\ref{fig:spatial}(b)) and found in
    literature\cite{kajtar_tropical_2017}. The absence of a sufficient number
    of long-range arcs makes hybrid and constraint-based algorithms incapable to
    model teleconnections and therefore unsuitable for propagating evidence.
  \item Tabu search and HC (with almost identical results) produce networks with
    the highest likelihood and the largest number of arcs for $\gamma < 10$
    (with $|A| > 2500$ for $\gamma \leq 0.2$). Even intermediate networks
    ($\gamma = 0.5$, $|A|= 2127$) include a large number of teleconnections and
    allow propagating evidence with realistic results (Figure
    \ref{fig:graphs}(b,d)).
  \item Score-based algorithms are faster than both hybrid and constraint-based
    algorithms. The difference in speed with \HHPC{} and MMHC for
    $\gamma \in \{0, 0.2, 0.5, 1, 1.5, 2\}$ is markedly larger, because in this
    range the score-based algorithms return DAGs containing more arcs than the
    hybrids for the same $\gamma$.
\end{itemize}

Finally, in Figure \ref{fig:results}(d) we examine the relative number of
unshielded v-structures in a network, defined as the number of unshielded
v-structures divided by the amount of adjacent pairs of arcs in a graph. In a DAG,
on average, 25\% of all adjacent pairs of arcs are (shielded or unshielded)
v-structures. The proportion of unshielded v-structures is smaller and depends
on $|N|$ and $|A|$.  For $N = 648$ and $|A| \in [25,7868]$, the average
proportion of unshielded v-structures over all possible DAGs lies between
$0.2499$ ($|A| = 25$) and $0.2125$ ($|A|=7868$). Note that, among the DAGs we
learned, those with up to 1500 arcs contain only short-range arcs and no
teleconnections. It is intuitive that most pairs of adjacent arcs connecting
nearby locations will not be modelled as an unshielded v-structure: they will be
part of a dense cluster of nodes that are dependent just because of local
weather patterns, and either the path is not a v-structure or the parents in the
v-structure are likely to be connected. For a dense DAG (with more than $1500$
arcs, as returned by HC and tabu search for $\gamma \leq 0.5$) it makes sense
that the amount of unshielded v-structures is higher than random as two nodes
corresponding to distant geographical locations will be connected by a path of
length two only when their association is strong enough to overcome the effect
of local weather patterns. Results in Figure \ref{fig:results}(d) show that all
algorithms seem to follow this intuition except for PC-Stable at large values of
$\gamma$ where it has the biggest relative amount of unshielded v-structures and
discovers more conditional dependence structure than random.

\subsection{Small Networks versus Large Networks (Climate Data)}

Different classes of structure learning algorithms learn networks with different
levels of sparsity when using ($\BICg$, $\GSQ_{\BICg}$). Since the number of
nodes in the networks is fixed by the geographical grid, we will treat sparse
graphs as ``small'' and dense graphs as ``large networks'' because the former
will have a smaller number of parameters and thus will represent simpler BNs.
All algorithms are able to learn small networks with up to $500$ arcs. Hybrid
and score-based algorithms can also learn medium networks with up to $1200$
arcs. Only score-based algorithms can successfully learn dense networks
containing up to $8000$ arcs. Constraint-based algorithms learn the most
accurate small networks in terms of log-likelihood.  Score-based algorithms
learn small networks faster than constraint-based algorithms and score-based
algorithms learn medium networks faster and more accurately than hybrid
algorithms. As score-based algorithms are the only algorithms that can model
large graphs, they are the only viable choice in that case. Since only large
graphs capture complex spatial dependencies we consider score-based algorithms
unique in their capacity to model climate data with short- and long-range
dependence structures.

\section{Discussion and Conclusions}

In this paper we revisited the problem of assessing different classes of BN
structure learning algorithms; we improved over existing comparisons of learning
accuracy and speed in the literature by removing the confounding effect of
different choices of statistical criteria. Interestingly, we found that
constraint-based algorithms are overall less accurate than tabu search (but not
simulated annealing) for both small and large sample sizes (\textbf{Q1}), but
are more accurate than other score-based algorithms in many simulation settings.
There is no systematic difference in accuracy between constraint-based and
hybrid algorithms (\textbf{Q3}). We also found that tabu search, as a
score-based algorithm, is often faster than most constraint-based and hybrid
algorithms (\textbf{Q2}). Finally, we found that hybrid algorithms are not
faster overall than constraint-based or score-based algorithms; in fact, there
was no consistent ordering of the algorithms from these classes across different
simulation scenarios (\textbf{Q4}). We noted that PC and RSMAX2 were
consistently among the fastest two constraint-based/hybrid algorithms for most
of the considered BNs and sample sizes. No systematic difference in the ranking
of different classes of algorithms in terms of speed and accuracy was observed
for any class of algorithms for small networks compared to large networks
(\textbf{Q5}).

All these conclusions are in contrast with other findings in the literature;
among others:
\begin{itemize}
  \item Tsamardinos \emph{et al.} \citep{mmhc} used a set of discrete reference
    BNs (including ALARM, CHILD, HAILFINDER and PIGS) to compare MMHC with tabu
    search, GES and PC (in its original formulation from \cite{spirtes}). They
    found MMHC to be faster than tabu search ($2.34\times$) and much faster than
    PC ($9.22\times$), while at the same time to have a smaller SHD
    ($1.85\times$ larger SHD for tabu search, $7.25\times$ for PC). However,
    these conclusions are limited by several issues: statistical criteria in
    different algorithms do not match; both BDeu's imaginary sample size and the
    significance threshold for the conditional independence tests are much
    larger than current best practices suggest \cite{ueno}; sample sizes in the
    simulation are absolute ($n$) instead of relative ($n/\Theta$), making the
    aggregation of the results problematic.
  \item Spirtes \cite{spirtes2} states that, unlike score-based algorithms,
    constraint-based algorithms ``are generally fast'', but that ``mistakes made
    early in constraint-based searches can lead to later mistakes'' which is
    exacerbated by ``the problem of multiple testing'' especially in large
    networks.
  \item Similarly, Koller and Friedman \cite{koller} state that constraint-based
    algorithms are ``sensitive to failures in individual independence tests''
    and that ``it suffices that one of these tests return a wrong answer to
    mislead the network construction procedure''; while score-based algorithms
    are ``less sensitive to individual failures'' but ``that they pose a
    search problem that may not have an elegant and efficient solution''.
  \item Natori \emph{et al.} \cite{natori} state that constraint-based
    algorithms can ``relax computational cost problems and can extend the
    available learning network size for learning'' compared to score-based
    algorithms. In the follow-up paper \cite{natori2}, where they compare the
    Recursive Autonomy Identification (RAI) \cite{rai} constraint-based
    algorithm with PC (in its original formulation) and MMHC using a a set of
    discrete reference BNs (including ALARM, ANDES, MUNIN and WIN95PTS), they
    confirmed this with a simulation study in PC and RAI scale better for large
    networks compared to MMHC. These results, however, are problematic because
    speed was measured in seconds and the simulations were run with bespoke
    implementations of the structure learning algorithms that were heterogeneous
    in terms of efficiency (Matlab vs Java). In addition, the table of results
    in \cite{natori2} is incomplete due to artificially limiting the running
    time of individual simulations.
  \item Niinim{\"a}ki and Parviainen \cite{sll} compare, among other algorithms,
    tabu search, GES and MMHC in terms of SHD and running time (in seconds) over
    $4$ discrete reference BNs (HAILFINDER and modified versions of ALARM,
    CHILD, INSURANCE). The figures included in the paper show MMHC as being both
    faster and more accurate than tabu search; and to be as accurate as GES
    while being faster. Again the results are limited by the confounding effect
    of choosing different statistical criteria, and by the measuring speed in
    absolute running times with heterogeneous software implementations.
\end{itemize}
In addition, we note that the literature referenced in the above list provides
these guidelines using only discrete BNs as a base, even when not stated
explicitly. Our conclusions about the relative speed and accuracy of various
classes of structure learning algorithms for GBNs is completely novel to the
best of our knowledge.

For complex data we found that only score-based algorithms produce large
networks in which higher-order dependencies are profoundly represented. In
climate data higher-order dependencies are related to teleconnections that are
key to model climate variability.

These results, which we confirmed on both simulated data and real-world complex
data, are intended to provide guidance for additional studies; we do not exclude
the existence of other sources of confounding, such as tuning parameters, which
should be further investigated.

\section*{Acknowledgements}

CEG and JMG were supported by the project MULTI-SDM (CGL2015-66583-R,
\linebreak MINECO/FEDER).

\bibliographystyle{elsarticle-num}
 \bibliography{biblio.bib}

\begin{thebibliography}{10}
\expandafter\ifx\csname url\endcsname\relax
  \def\url#1{\texttt{#1}}\fi
\expandafter\ifx\csname urlprefix\endcsname\relax\def\urlprefix{URL }\fi
\expandafter\ifx\csname href\endcsname\relax
  \def\href#1#2{#2} \def\path#1{#1}\fi

\bibitem{cowell}
R.~Cowell, {Conditions Under Which Conditional Independence and Scoring Methods
  Lead to Identical Selection of Bayesian Network Models}, in: {Proceedings of
  the 17th Conference on Uncertainty in Artificial Intelligence}, 2001, pp.
  91--97.

\bibitem{koller}
D.~Koller, N.~Friedman, {Probabilistic Graphical Models: Principles and
  Techniques}, MIT Press, 2009.

\bibitem{chickering}
D.~M. Chickering, {A Transformational Characterization of Equivalent Bayesian
  Network Structures}, in: P.~Besnard, S.~Hanks (Eds.), {Proceedings of the
  11th Conference on Uncertainty in Artificial Intelligence}, Morgan Kaufmann,
  1995, pp. 87--98.

\bibitem{heckerman}
D.~Heckerman, D.~Geiger, D.~M. Chickering, {Learning Bayesian Networks: The
  Combination of Knowledge and Statistical Data}, Machine Learning 20~(3)
  (1995) 197--243.

\bibitem{heckerman3}
D.~Geiger, D.~Heckerman, {Learning Gaussian Networks}, in: {Proceedings of the
  10th Conference on Uncertainty in Artificial Intelligence}, 1994, pp.
  235--243.

\bibitem{weatherburn}
C.~E. Weatherburn, {A First Course in Mathematical Statistics}, Cambridge
  University Press, 1961.

\bibitem{copula}
G.~Elidan, {Copula Bayesian Networks}, in: J.~D. Lafferty, C.~K.~I. Williams,
  J.~Shawe-Taylor, R.~S. Zemel, A.~Culotta (Eds.), Advances in Neural
  Information Processing Systems 23, 2010, pp. 559--567.

\bibitem{truncexp}
S.~Moral, R.~Rumi, A.~Salmer{\'o}n, {Mixtures of Truncated Exponentials in
  Hybrid Bayesian Networks}, in: {Symbolic and Quantitative Approaches to
  Reasoning with Uncertainty (ECSQARU)}, Vol. 2143 of Lecture Notes in Computer
  Science, Springer, 2001, pp. 156--167.

\bibitem{lauritzen}
S.~L. Lauritzen, N.~Wermuth, {Graphical Models for Associations Between
  Variables, Some of which are Qualitative and Some Quantitative}, The Annals
  of Statistics 17~(1) (1989) 31--57.

\bibitem{npcomp}
D.~M. Chickering, {Learning Bayesian networks is NP-Complete}, in: D.~Fisher,
  H.~Lenz (Eds.), Learning from Data: Artificial Intelligence and Statistics V,
  Springer-Verlag, 1996, pp. 121--130.

\bibitem{nplarge}
D.~M. Chickering, D.~Heckerman, C.~Meek, {Large-sample Learning of Bayesian
  Networks is NP-hard}, Journal of Machine Learning Research 5 (2004)
  1287--1330.

\bibitem{notnphard}
T.~Claassen, J.~M. Mooij, T.~Heskes, {Learning Sparse Causal Models is not
  NP-hard}, in: {Proceedings of the 29th Conference on Uncertainty in
  Artificial Intelligence}, 2013, pp. 172--181.

\bibitem{schwarz}
G.~Schwarz, {Estimating the Dimension of a Model}, The Annals of Statistics
  6~(2) (1978) 461--464.

\bibitem{harary}
F.~Harary, E.~M. Palmer, {Graphical Enumeration}, Academic Press, 1973.

\bibitem{csprior}
R.~Castelo, A.~Siebes, {Priors on Network Structures. Biasing the Search for
  Bayesian Networks}, International Journal of Approximate Reasoning 24~(1)
  (2000) 39--57.

\bibitem{mukherjee}
S.~Mukherjee, T.~P. Speed, {Network Inference Using Informative Priors},
  Proceedings of the National Academy of Sciences 105~(38) (2008) 14313--14318.

\bibitem{k2}
G.~F. Cooper, E.~Herskovits, {A Bayesian Method for Constructing Bayesian
  Belief Networks from Databases}, in: {Proceedings of the 7th Conference on
  Uncertainty in Artificial Intelligence}, 1991, pp. 86--94.

\bibitem{ic}
T.~S. Verma, J.~Pearl, {Equivalence and Synthesis of Causal Models},
  Uncertainty in Artificial Intelligence 6 (1991) 255--268.

\bibitem{spirtes}
P.~Spirtes, C.~Glymour, R.~Scheines, {Causation, Prediction, and Search}, MIT
  Press, 2000.

\bibitem{colombo}
D.~Colombo, M.~H. Maathuis, {Order-Independent Constraint-Based Causal
  Structure Learning}, Journal of Machine Learning Research 15 (2014)
  3921--3962.

\bibitem{mphd}
D.~Margaritis, {Learning Bayesian Network Model Structure from Data}, Ph.D.
  thesis, School of Computer Science, Carnegie-Mellon University, Pittsburgh,
  PA (May 2003).

\bibitem{fastiamb2}
S.~Yaramakala, D.~Margaritis, {Speculative Markov Blanket Discovery for Optimal
  Feature Selection}, in: {ICDM '05: Proceedings of the Fifth IEEE
  International Conference on Data Mining}, IEEE Computer Society, 2005, pp.
  809--812.

\bibitem{hitonpc}
C.~F. Aliferis, A.~Statnikov, I.~Tsamardinos, S.~Mani, X.~D. Xenofon, {Local
  Causal and Markov Blanket Induction for Causal Discovery and Feature
  Selection for Classification Part I: Algorithms and Empirical Evaluation},
  Journal of Machine Learning Research 11 (2010) 171--234.

\bibitem{bouckaert}
R.~R. Bouckaert, {Bayesian Belief Networks: from Construction to Inference},
  Ph.D. thesis, Utrecht University, The Netherlands (1995).

\bibitem{larranaga}
P.~Larra{\~n}aga, B.~Sierra, M.~J. Gallego, M.~J. Michelena, J.~M. Picaza,
  {Learning Bayesian Networks by Genetic Algorithms: A Case Study in the
  Prediction of Survival in Malignant Skin Melanoma}, in: {Proceedings of the
  6th Conference on Artificial Intelligence in Medicine in Europe (AIME'97)},
  Springer, 1997, pp. 261--272.

\bibitem{norvig}
S.~J. Russell, P.~Norvig, {Artificial Intelligence: A Modern Approach}, 3rd
  Edition, Prentice Hall, 2009.

\bibitem{ges}
D.~M. Chickering, {Optimal Structure Identification With Greedy Search},
  Journal of Machine Learning Research 3 (2002) 507--554.

\bibitem{cutting}
J.~Cussens, {Bayesian Network Learning with Cutting Planes}, in: {Proceedings
  of the 27th Conference on Uncertainty in Artificial Intelligence}, 2012, pp.
  153--160.

\bibitem{suzuki17}
J.~Suzuki, {An Efficient Bayesian Network Structure Learning Strategy}, New
  Generation Computing 35~(1) (2017) 105--124.

\bibitem{scanagatta}
M.~Scanagatta, C.~P. de~Campos, G.~Corani, M.~Zaffalon, {Learning Bayesian
  Networks with Thousands of Variables}, in: Advances in Neural Information
  Processing Systems 28, 2015, pp. 1864--1872.

\bibitem{madigan-york}
D.~Madigan, J.~York, {Bayesian Graphical Models for Discrete Data},
  International Statistical Review 63 (1995) 215--232.

\bibitem{husmeier-mcmc}
M.~Grzegorczyk, D.~Husmeier, {Improving the Structure MCMC Sampler for Bayesian
  Networks by Introducing a New Edge Reversal Move}, Machine Learning 71 (2008)
  265--305.

\bibitem{kuipers-mcmc}
J.~Kuipers, G.~Moffa, {Partition MCMC for Inference on Acyclic Digraphs},
  Journal of the American Statistical Association 112~(517) (2017) 282--299.

\bibitem{order1}
P.~Larra{\~n}aga, C.~M.~H. Kuijpers, R.~H. Murga, Y.~Yurramendi, {Learning
  Bayesian Network Structures by Searching for the Best Ordering with Genetic
  Algorithms}, {IEEE Transactions on Systems, Man, and Cybernetics - Part A:
  Systems and Humans} 26~(4) (1996) 487--493.

\bibitem{corani}
M.~Scanagatta, G.~Corani, M.~Zaffalon, {Improved Local Search in Bayesian
  Networks Structure Learning}, Proceedings of Machine Learning Research (AMBN
  2017) 73 (2017) 45--56.

\bibitem{catnet}
N.~Balov, P.~Salzman, catnet: Categorical Bayesian Network Inference, r package
  version 1.15.3 (2017).

\bibitem{mmhc}
I.~Tsamardinos, L.~E. Brown, C.~F. Aliferis, {The Max-Min Hill-Climbing
  Bayesian Network Structure Learning Algorithm}, Machine Learning 65~(1)
  (2006) 31--78.

\bibitem{magic14}
M.~Scutari, P.~Howell, D.~J. Balding, I.~Mackay, {Multiple Quantitative Trait
  Analysis Using Bayesian Networks}, Genetics 198~(1) (2014) 129--137.

\bibitem{h2pc}
M.~Gasse, A.~Aussem, H.~Elghazel, {A Hybrid Algorithm for Bayesian Network
  Structure Learning with Application to Multi-Label Learning}, Expert Systems
  with Applications 41~(15) (2014) 6755--6772.

\bibitem{agresti}
A.~Agresti, Categorical Data Analysis, 3rd Edition, Wiley, 2012.

\bibitem{ueno}
M.~Ueno, {Learning Networks Determined by the Ratio of Prior and Data}, in:
  {Proceedings of the 26th Conference on Uncertainty in Artificial
  Intelligence}, 2010, pp. 598--605.

\bibitem{suzuki16}
J.~Suzuki, {A Theoretical Analysis of the BDeu Scores in Bayesian Network
  Structure Learning}, Behaviormetrika 44 (2016) 97--116.

\bibitem{pgm16}
M.~Scutari, {An Empirical-Bayes Score for Discrete Bayesian Networks}, Journal
  of Machine Learning Research (Proceedings Track, PGM 2016) 52 (2016)
  438--448.

\bibitem{behaviormetrika18}
M.~Scutari, {Dirichlet Bayesian Network Scores and the Maximum Relative Entropy
  Principle}, Behaviormetrika 45~(2) (2018) 337--362.

\bibitem{kuipers}
J.~Kuipers, G.~Moffa, D.~Heckerman, {Addendum on the Scoring of Gaussian
  Directed Acyclic Graphical Models}, The Annals of Statistics 42~(4) (2014)
  1689--1691.

\bibitem{cugnata}
F.~Cugnata, R.~S. Kenett, S.~Salini, {Bayesian Networks in Survey Data:
  Robustness and Sensitivity Issues}, Journal of Quality Technology 4~(3)
  (2016) 253--264.

\bibitem{spirtes2}
P.~Spirtes, {Introduction to Causal Inference}, Journal of Machine Learning
  Research 11 (2010) 1643--1662.

\bibitem{natori}
K.~Natori, M.~Uto, Y.~Nishiyama, S.~K.~M. Ueno, {Constraint-Based Learning
  Bayesian Networks Using Bayes Factor}, in: {Advanced Methodologies for
  Bayesian Networks}, Springer, 2015, pp. 15--31.

\bibitem{bnrepo}
M.~Scutari, {Bayesian Network Repository}, http://www.bnlearn.com/bnrepository
  (2012).

\bibitem{jss09}
M.~Scutari, {Learning Bayesian Networks with the bnlearn R Package}, Journal of
  Statistical Software 35~(3) (2010) 1--22.

\bibitem{tetrad}
J.~A. Landsheer, {The Specification of Causal Models with Tetrad IV: A Review},
  Structural Equation Modeling 17~(4) (2010) 703--711.

\bibitem{r-causal}
C.~Wongchokprasitti, {rcausal: R-Causal Library}, r package version 0.99.9
  (2017).

\bibitem{fountalis_spatio-temporal_2014}
I.~Fountalis, A.~Bracco, C.~Dovrolis, {Spatio-Temporal Network Analysis for
  Studying Climate Patterns}, Climate Dynamics 42~(3-4) (2014) 879--899.

\bibitem{tsonis_role_2008}
A.~A. Tsonis, K.~L. Swanson, G.~Wang, {On the Role of Atmospheric
  Teleconnections in Climate}, Journal of Climate 21~(12) (2008) 2990--3001.

\bibitem{PhysRevLett.100.228501}
K.~Yamasaki, A.~Gozolchiani, S.~Havlin, {Climate Networks around the Globe are
  Significantly Affected by El Ni\~no}, Phys. Rev. Lett. 100 (2008) 228501.

\bibitem{tsonis_what_2006}
A.~A. Tsonis, K.~L. Swanson, P.~J. Roebber, {What Do Networks Have to Do with
  Climate?}, Bulletin of the American Meteorological Society 87~(5) (2006)
  585--595.

\bibitem{cano}
R.~Cano, C.~Sordo, J.~M. Guti{\'e}rrez, {Applications of Bayesian Networks in
  Meteorology}, in: J.~A. G{\'a}mez, S.~Moral, A.~Salmer{\'o}n (Eds.), Advances
  in {Bayesian} {Networks}, Springer, 2004, pp. 309--328.

\bibitem{10.2307/24310025}
J.~Chen, Z.~Chen, {Extended BIC For Small-n-Large-p Sparse GLM}, Statistica
  Sinica 22~(2) (2012) 555--574.

\bibitem{kajtar_tropical_2017}
J.~B. Kajtar, A.~Santoso, M.~H. England, W.~Cai, {Tropical Climate Variability:
  Interactions Across the Pacific, Indian, and Atlantic Oceans}, Climate
  Dynamics 48~(7--8) (2017) 2173--2190.

\bibitem{natori2}
K.~Natori, M.~Uto, M.~Ueno, {Consistent Learning Bayesian Networks with
  Thousands of Variables}, Proceedings of Machine Learning Research (AMBN 2017)
  73 (2017) 57--68.

\bibitem{rai}
R.~Yehezkel, B.~Lerner, {Bayesian Network Structure Learning by Recursive
  Autonomy Identification}, Journal of Machine Learning Research 10 (2009)
  1527--1570.

\bibitem{sll}
T.~Niinim{\"a}ki, P.~Parviainen, {Local Structure Discovery in Bayesian
  Networks}, in: {Proceedings of the 28th Conference on Uncertainty in
  Artificial Intelligence}, 2012, pp. 634--643.

\end{thebibliography}

\end{document}